\documentclass[12pt,a4paper,jmp,nofootinbib]{revtex4-1}

\usepackage{graphicx}
\usepackage{amsmath,amsthm}
\usepackage{amssymb,amsfonts}
\usepackage{bbold}
\usepackage{hyperref}
\usepackage{xargs}
\usepackage{mathtools}
\usepackage{tikz}
\usepackage{tikz-cd}
\usepackage{mathrsfs}
\usepackage{eucal}
\usepackage{physics}
\usepackage{amsthm}

\usetikzlibrary{arrows.meta} 
\usetikzlibrary{calc}

\usepackage[margin=2.5cm]{geometry}

\def\nfoot#1{{\begingroup\linespread{0.75}\footnote{#1}\endgroup}}

\newcommand{\fixedwidth}[2]{%
  \makebox[\widthof{$#1$}][l]{\vphantom{$#1$}\smash{$#2$}}%
}

\def\physbox#1{\text{\fbox{ \ensuremath{#1} }}}

\newcommand{\R}{\mathbb{R}}
\newcommand{\C}{\mathbb{C}}
\newcommand{\bH}{\mathbb{H}}
\renewcommand{\P}{\mathbb{P}}
\newcommand{\Z}{\mathbb{Z}}

\newcommand{\N}{\mathcal{N}}

\DeclareMathOperator{\im}{im}

\DeclareMathOperator{\coker}{coker}
\DeclareMathOperator{\CE}{CE}

\def\defeq{\doteq}

\usepackage{xparse}
\NewDocumentCommand \deq { o m }{
\begin{equation}
#2
\IfNoValueF{#1}{\label{#1}}
\end{equation}
}

\def\lie#1{\mathfrak{#1}}
\DeclareMathOperator{\Spin}{Spin}

\DeclareMathOperator{\Cl}{Cl}
\DeclareMathOperator{\Fl}{Flag}

\DeclareMathOperator{\End}{End}
\DeclareMathOperator{\Sym}{Sym}
\DeclareMathOperator{\Ann}{Ann}

\DeclareMathOperator{\OG}{OGr}
\DeclareMathOperator{\Gr}{Gr}
\DeclareMathOperator{\PS}{PS}

\DeclareMathOperator{\Fun}{Fun}

\DeclareMathOperator{\Spec}{Spec}
\DeclareMathOperator{\id}{id}
\def\Nilpp(#1,#2){Y\qty({#1};{#2})}
\def\Nilp(#1,#2){\widehat{Y}\qty({#1};{#2})}
\def\OO{\mathscr{O}}

\def\qu(#1){\ensuremath{\lambda^{#1}}}
\def\ttensor{\mathrel{\tilde{\otimes}}}

\def\Hilb{\mathscr{H}}

\def\Berk{\mathscr{D}}
\def\PSSF(#1,#2){{\Xi\qty({#1};{#2})}}
\def\XPSSF(#1,#2){{\widehat{\Xi}\qty({#1};{#2})}}

\DeclareMathOperator{\ad}{ad}

\def\rep#1{\mathbf{#1}}
\def\crep#1{\smash{\mathbf{\overline{#1}}}}

\usepackage{stmaryrd}

\numberwithin{satz}{section}

\theoremstyle{definition}
\newtheorem*{definition}{Definition}

\numberwithin{subsubsection}{subsection}
\numberwithin{subsection}{section}
\numberwithin{equation}{section}

\makeatletter
\def\p@subsection{}
\def\p@subsubsection{}
\makeatother

\newcommand{\nocontentsline}[3]{}
\newcommand{\tocless}[2]{\bgroup\let\addcontentsline=\nocontentsline#1{#2}\egroup}

\begin{document}

\title{Nilpotence varieties}

\author{Richard Eager}
\email{eager@mathi.uni-heidelberg.de}
\author{Ingmar Saberi}
\email{saberi@mathi.uni-heidelberg.de}
\author{Johannes Walcher}
\email{walcher@uni-heidelberg.de}
\affiliation{Mathematisches Institut der Ruprecht-Karls-Universit\"at Heidelberg \\ Im Neuenheimer Feld 205 \\ 69120 Heidelberg, Germany}

\begin{abstract}
We consider algebraic varieties canonically associated to any Lie superalgebra, and study them in detail for super-Poincar{\'e} algebras of physical interest. They are the locus of nilpotent elements in (the projectivized parity reversal of) the odd part of the algebra. Most of these varieties have appeared in various guises in previous literature, but we study them systematically here, from a new perspective:
as the natural moduli spaces parameterizing twists of a super-Poincar{\'e}-invariant physical theory. We obtain a classification of all possible twists, as well as a systematic analysis of unbroken symmetry in twisted theories. The natural stratification of the varieties, the identification of strata with twists, and the action of Lorentz and $R$-symmetry on the varieties are emphasized. We also include a short and unconventional exposition of the pure-spinor superfield formalism, from the perspective of twisting, and demonstrate that it can be applied to construct familiar multiplets in four-dimensional minimally supersymmetric theories; in all dimensions and with any amount of supersymmetry, this technique produces BRST or BV complexes of supersymmetric theories from the Koszul complex of the cone point over the coordinate ring of the nilpotence variety, possibly tensored with a module over that coordinate ring. In addition, we remark on a natural emergence of nilpotence varieties in the Chevalley--Eilenberg cohomology of supertranslations, and give two applications related to these ideas: a calculation of Chevalley--Eilenberg cohomology for the six-dimensional $\N=(2,0)$ supertranslation algebra, and
 a BV complex matching the field content of type IIB supergravity from the coordinate ring of the corresponding nilpotence variety.
\end{abstract}

\maketitle

\addtocounter{page}{-1}

\setcounter{tocdepth}{2}
\tableofcontents

\tocless\bibsection


\section{General philosophy}
\label{sec:phil}

In this note, we study a general family of algebraic varieties associated to super Lie algebras. Such an algebra, as the reader will recall, takes the general form
\deq[eq:superLie]{
A = A^0 \oplus A^1,
}
where the superscript refers to a grading by~$\Z/2\Z$ that determines the parity of the bracket according to the Koszul sign rule. Such algebras were originally defined in~\cite{BerezinKac}, and have been the subject of intense study since then.

Expanding out the bracket into its graded components, the adjoint action of $A$ on itself implies that $A^0$ is an ordinary Lie algebra, $A^1$ carries a representation of~$A^0$, and the anticommutator map 
\deq[eq:oddbracket]{
\{\cdot,\cdot\}: \Sym^2(A^1) \rightarrow A^0
}
is an intertwiner of $A^0$-representations. 

Let $Q$ denote an element of~$A^1$, and consider the equations
\deq{
Q^2 \defeq \frac 12\{Q,Q\} = 0.
}
(This consists of several equations, since the left-hand side is \emph{a~priori} an element of~$A^0$.) Together, they form a set of homogeneous quadrics in the vector space~$A^1$; their space of solutions thus descends to define a projective variety
\deq{
Y \subset P(A^1)
}
over the ground field~$k$ of the algebra~$A$.
We will denote the corresponding affine variety by~$\widehat{Y} \subset A^1$. Such a projective variety $Y$ is associated to any super Lie algebra; we call it the \emph{nilpotence variety} of~$A$. 

Both definitions and many examples of such varieties have appeared in the literature before. The general study of the variety of nilpotent elements was introduced by~\cite{Gruson}, and used to prove various results about the cohomology of Lie superalgebras; her definitions were later generalized in~\cite{DufloSerganova} to include the ``associated variety'' of an $A$-module, which is a certain subvariety of~$Y(A)$ respecting its stratification. In those works, however, the emphasis was on simple or semi-simple super-Lie algebras, rather than super-Poincar{\'e} algebras.

Independently, in the physics literature, some of the varieties~$Y$ for super-{Poincar\'e} algebras have been studied in the context of the pure spinor formalism.\nfoot{In the literature, $Y$ is often called a ``pure spinor space;'' we opt to reserve this term for the space of pure spinors as defined by Cartan~\cite{Cartan}. See below~(\S\ref{ssec:OG}).}  
The most famous example is the variety associated to the ten-dimensional minimal supersymmetry algebra, which appears in the Berkovits pure spinor formalism~\cite{BerkovitsPS,Berkovits-coho}. 
Other examples have been studied for the purposes of generalizing the pure spinor formalism in~\cite{Cederwall-coho,Cederwall-6d,Movshev:2011cy,MovshevSchwarz,Movshev-YM} (just for example). Independently, and for different reasons, some very simple examples for theories with four supercharges were written down in~\cite{GNSSS}.
We study all instances here systematically, though, and from a new perspective, which relates the varieties to one another and to the procedure of twisting a supersymmetric theory. As an additional benefit, this allows us to clarify the relationship between twisting and the pure spinor formalism.

The primary physics motivation for studying superalgebras is interest in supersymmetric, (locally) Lorentz-invariant field theories. Such a field theory admits the action of an algebra $A$ taking a particular form. 
For non-conformally invariant theories, 
the relevant algebras are the 
(extended) super-Poincar\'e algebras; this means that 
\deq[eq:supP]{
A^0 = \qty( V \rtimes \lie{so}(V) ) \times R,
}
where $V\cong\R^n$ is the inner product space pointwise isomorphic to the tangent space of 
the space-time manifold (before complexification), $\lie{so}(V)$ is the Lie algebra of the bilinear form on $V$,
and $R$ is some other ordinary Lie algebra with a non-degenerate invariant bilinear form.

One also demands that $A^1$ sits in a spinorial representation of~$\lie{so}(V)$ (or possibly consists of  several copies of irreducible spin representations), 
and that the map~\eqref{eq:oddbracket} 
arises from the standard bilinear mappings from spinors to the vector representation, $V$,
as well as possibly to the trivial representation (in the presence of ``central charges'').

In fact, the Coleman--Mandula theorem~\cite{WessBagger} 
implies that the only finite-dimensional physically relevant possibilities for the algebra~$A^0$ 
are those of the form~\eqref{eq:supP},
conformal algebras (which are of the form $\lie{so}(d+1,1)$ in Euclidean dimension~$d$), and 
infinite-dimensional algebras, such as the Virasoro and Kac--Moody algebras, relevant to 
two-dimensional models. 
When one allows super Lie algebras, the first of these possibilities may be extended (according to Haag--{\L}opusza{\'n}ski--Sohnius~\cite{WessBagger})
by a spinorial representation in odd degree, as in \eqref{eq:oddbracket}.
The second extends to those super Lie algebras 
appearing in Nahm's list of superconformal algebras~\cite{Nahm}, and there are correspondingly infinite-dimensional superconformal algebras relevant to two-dimensional physics.
Only the first possibility will be of 
concern in this article, although it would be interesting to repeat exactly the same set 
of exercises for the superalgebras appearing on Nahm's list. We leave that task to future 
work; a study of the variety~$Y$ relevant to the four-dimensional $\N=2$ superconformal 
algebra will appear in~\cite{GNPS-forthcoming}.

In the real physical theory, the inner product on $V$ has Minkowski signature, $\lie{so}(V)$ 
is the Lie algebra of the Lorentz group, and $R$ generates a 
compact Lie group via its unitary representation on a complex Hilbert space,
$\Hilb$. As a consequence of the positivity of the inner product on~$\Hilb$, the kernel of the 
map~\eqref{eq:oddbracket} would be represented trivially on $\Hilb$, so that it
is no loss to assume that the map is non-degenerate. In studying the
representation theory of \eqref{eq:superLie}, it is also natural to
view scalars as extended to $\C$, 
and we will
work with the complexified version of \eqref{eq:superLie} throughout. We
will denote the (henceforth complex) dimension of $V$ by $d$, and sometimes
write $\lie{so}(d)=\lie{so}(d,\C)$ for the associated Lie algebra for clarity.

While the letter $R$ is meant to suggest $R$-symmetry, we leave open the possibility 
that $R$ includes flavor symmetries and central charges; this will have no effect on 
the following discussion. By definition, a symmetry is termed an $R$-symmetry if it 
is represented nontrivially on~$A^1$, and a flavor symmetry otherwise. Since $R$ 
commutes with $\lie{so}(V)$, Schur's lemma allows us to write
\deq[eq:decomp]{
A^1= S\ttensor U,
}
where the symbol denotes some direct sum of spinor representations~$S$ tensored with auxiliary vector spaces~$U$ carrying representations of~$R$.  
We recall the standard pattern below in~\S\ref{sec:spinors}.
Suffice it say for now that the ``amount of supersymmetry'' $\N$ is either a single integer 
or a pair of such integers
when
$d=2\bmod 4$. $U$ also carries a (symmetric or anti-symmetric) non-degenerate
inner product to accompany the map $S\otimes S\rightarrow V$ in the definition of~\eqref{eq:oddbracket}; $R$ is, at least in the simplest cases, the subalgebra of~$\lie{gl}(U)$ preserving this pairing.

Such super-Poincar\'e algebras can be constructed in various dimensions; however, 
due to physical considerations 
regarding the existence of representations that can be used to construct sensible field 
theories, one never considers cases where $\dim(A^1)>16$, implying in particular that
$d\le 10$. The only exceptions are the algebras relevant to eleven-dimensional supergravity, 
for which $d=11$ and $\dim(A^1) = 32$, and those of type II supergravity theories in ten 
dimensions; these theories, though, automatically include dynamical metric degrees of freedom.

Supersymmetric theories admit various \emph{twists;} the term refers to a standard procedure for extracting a subsector of the theory, which depends topologically or possibly holomorphically on the spacetime. Twisted theories have been of great interest, not only in physics, but also in mathematics (geometry and topology in particular). In the standard description one finds in the literature of this procedure, one tends to see a two-part story along the following lines: One would like to take cohomology of a supercharge, but (since supercharges are always in the spinor representation of the Lorentz group) one cannot. Therefore, one modifies the action of the Lorentz group according to an embedding 
\begin{equation}
\label{eq:twisthomo}
\begin{tikzcd}
\lie{so}(d)'  \arrow{r}{1\times\phi} & \lie{so}(d) \times R \subseteq A^0.
\end{tikzcd}
\end{equation}
$\phi$ is thought of as a ``twisting homomorphism.'' Having made this choice, it may be that $A^1$ contains one or more trivial representations of~$\lie{so}(d)'$; such fermionic symmetries are then necessarily nilpotent, and one can pass to their cohomology to obtain a topological field theory.

It is perhaps slightly unfortunate that the word ``twist'' for this procedure is so thoroughly established in the mathematical physics literature. The word would be overburdened even without this usage, and the problem is only compounded by the fact that the piece of the apparatus that warrants the name ``twisting''  (the homomorphism $\phi$) can in fact often be viewed as a consequence of the procedure, rather than an independent choice. 
One of the central philosophical points of this note is to take the perspective that the fundamental data of the twist is a choice of nilpotent $Q\in A^1$; once this data is given, all else (including the set of possible twist maps---$\phi$ need not be unique, but often is) in fact follows.

For us, then, to ``twist'' a theory means to take the invariants of a fermionic symmetry. This procedure can be performed whenever the chosen symmetry operator~$Q$ is nilpotent; the invariants are then the cohomology of that operator (which can be thought of as all invariants, modulo those that are invariant for an uninteresting reason, and in fact belong to multiplets where $Q$ is represented nontrivially). The set of possible such symmetry operators is thus the variety~$\Nilpp(d,\N)$. In our notation, the pair~$(d;\N)$ follows the standard physics convention of labeling the algebra $A$ by the dimension and the integer (or pair of integers) giving the ``amount of extended supersymmetry'' as recalled before. 

This means that the collection of all possible twists of a theory with $(d;\N)$~supersymmetry fit together, in a sense, into a \emph{natural family} over the space~$\Nilpp(d,\N)$. Indeed, by virtue of its construction, $\Nilpp(d,\N)$ carries many interesting canonical bundles. The most immediate of these is a canonical line bundle $\mathscr{L}$ of nilpotent operators, spanned by~$Q$ over the point~$Q$:
\begin{equation}
\begin{tikzcd}[column sep = 1 em]
\C Q \ar[r] \ar[d] & \mathscr{L} \ar[rr] \ar[dr] & & A \times \Nilpp(d,\N) \ar[dl]  \\
Q \arrow[hook]{rr}{i} & & \Nilpp(d,\N) &
\end{tikzcd}
\end{equation}
(Here, the leftmost square is a pullback diagram over the inclusion $i$ of a point into~$\Nilpp(d,\N)$, identifying~$\C Q$ as the corresponding fiber of~$\mathscr{L}$.)

Thus, given any $A$-module $M$, one can form a natural bundle of chain complexes over~$\Nilpp(d,\N)$ by a sort of associated bundle or ``spreading out'' construction. Concretely, one takes the trivial bundle $M\times \Nilpp(d,\N)$ of~$A$-modules, which one thinks of as acted on at each point by a copy of~$A$. This then defines a bundle of chain complexes by pulling back to the action of~$\mathscr{L} \subset A\times Y(d;\N)$, which exhibits the action of~$\C[Q]/Q^2$ (that being, of course, equivalent to a differential). When $M$ is (for example) the collection of local operators in the untwisted theory, the fiber over a point consists of the local operators in the corresponding twist of the theory. For a different, particular choice of~$M$---an unconstrained superfield---a global version of the construction we have sketched here (the pure spinor superfield formalism of the physics literature) produces familiar supermultiplets and even BV complexes of familiar theories.  We give a more detailed exposition of this technique below in~\S\ref{sec:PSSF}.

There are two important and rather violent consequences of restricting the observables to $Q$-invariants of the theory. First of all, a subset of the bosonic symmetries in fact become $Q$-exact, and can be viewed as ``pure gauge'' in that sense. Specifically, let us define
\deq[eq:Edef]{
E = \{Q,A^1\}. 
}
For a super-Poincar\'e algebra, $E$ will contain a non-trivial subspace of the translation generators~$V$, and, by standard arguments, correlation functions of $Q$-closed operators will be independent of translations generated in that subspace. In a local theory, in which charges are integrated currents, the translations come from an exact energy-momentum tensor, and the correlation functions will be independent of deformations of certain components of the metric.

On the other hand, the symmetry algebra will also be broken to the commutant of~$Q$, which is a subalgebra~$Z(Q)\subseteq A$. (Other symmetries will not preserve $Q$, and therefore do not act on $Q$-cohomology.)
In fact, to be precise, a larger algebra is left unbroken, the \emph{stabilizer} of the line spanned by~$Q$:
\deq[eq:idealdef]{
I(Q) = \{ x \in A : [x,Q] \propto Q \}.
}
It follows from the Jacobi identity that both $Z(Q)$ and~$I(Q)$ are closed under the bracket.
Moreover, it is clear that $Z(Q) \subseteq I(Q)$, and that $E\subset Z(Q)$.%
\nfoot{Indeed, for super-Poincar\'e algebras, all of~$V$ is in~$Z(Q)$; $Z(Q)$ contains~$E$, however, for arbitrary super Lie algebras. In superconformal algebras, a novel feature is that the choice of~$Q$ may break some translations. As an immediate consequence, $Q$-cohomology in superconformal theories may not preserve homogeneity or isotropy; operators may acquire nontrivial support conditions. For an example of such a construction in recent literature, see~\cite{Rastelli}.}
And it is not difficult to see that $[I(Q),I(Q)]\subset Z(Q)$,
so that $I(Q)/Z(Q)$ is abelian.
The difference between the two can thus be summed up by saying that $I(Q)$ may contain additional generators of~$U(1)$ symmetries, with respect 
to which $Q$ transforms with definite (but nonzero) charge. $I(Q)$ is therefore the relevant
algebra to consider for questions related to gradings.
For similar reasons as above, both of these also define natural families of algebras over~$\Nilpp(d,\N)$. (The bracket above is the bracket of~$A$, 
and so is to be understood as either a commutator or anticommutator, as appropriate to the 
parity of~$x$.) 

With respect to either of the algebras $Z(Q)$ and~$I(Q)$, $Q$ is tautologically a scalar, perhaps carrying some definite $U(1)$ charges. For general reasons, though, $I(Q)$ can contain neither the generators of Lorentz nor any semisimple $R$-symmetry (in the first case, since $A^1$ must be a spinor, and in the second case by definition). Nonetheless, $I(Q)_\R$ may contain a subalgebra $\mathfrak{so}(d)'$, isomorphic to the Lorentz algebra, but defined by the graph of some nontrivial homomorphism $\phi$ from the Lorentz group to the $R$-symmetry group, as in \eqref{eq:twisthomo}.
When such a ``twisted'' Lorentz symmetry is unbroken, and additionally $E = V$, one obtains a ``topological theory'' by 
taking the Lorentz group to act on fields by $\mathfrak{so}(d)'$,
in addition to taking $Q$-invariants: In the twisted theory, the correlation functions (of $Q$-cohomology classes) are independent of position. The theory can be defined on any Riemannian manifold, but its observables are also independent of deformations of the metric.
Note that for any given nilpotent $Q$, there might be several possible twisting maps. Moreover, given the twisting map, there might be several different nilpotent scalar $Q$'s.

More generally, though, the Lorentz algebra may simply be broken to some subalgebra. Twists of this kind have recently attracted greater attention in the literature; see, just for example,~\cite{CostelloScheimbauer,CostelloLi,CostelloYangian}. In the case of a ``holomorphic'' twist, when $d=2n$ is even, the unbroken (real Lorentz) subgroup contains a factor isomorphic to 
$U(n)\subseteq SO(2n,\R)$, the subgroup preserving a particular complex structure. Since the branching rules for the spinor under $U(n)$ always contains a scalar, this class of twists will occur even in theories with minimal supersymmetry.
There may even be twists which are intermediate between holomorphic and topological, for which $E\subset V$ is a coisotropic subspace that is neither maximal nor minimal; an example was studied in~\cite{Kapustin-halftwist}. Additionally, it may happen that $E = V$, but (in spite of this) $Z^0(Q)_\R$ contains no $\lie{so}(d)'$ subalgebra. Examples will occur, for example, in dimensions eight and seven; the corresponding twists were first discussed by~\cite{BaulieuKannoSinger} and~\cite{Acharya:1997jn}, respectively.  In this case, one can define a topological theory that nonetheless only admits a formulation on manifolds of reduced holonomy. We will call such theories weakly topological.

This means that there is an interesting stratification of~$\Nilpp(d,\N)$, by what type of twist a given operator generates---or, put differently, by the form of the stabilizer algebra $I(Q)$. This stratification is precisely related to the jump loci of $Q$-cohomology in the family of chain complexes from above, and also to the decomposition of the variety into orbits of~$\lie{so}(d)\times R$. We will see that the structure of these varieties contains a lot of information, both about the structure that is preserved by a given twist, and about how various twists sit in relation to one another. In particular, thinking about the nilpotence variety makes it clear when a ``more topological'' twist can be obtained by deforming away from a ``less topological'' one, giving rise to spectral sequences relating the operators of the corresponding twisted theories in cases where both are defined~\cite{GNSSS}.

The local structure of $\Nilpp(d,\N)$ and its affine version can 
also be understood purely by studying the super-Poincar\'e algebra,
in particular by analyzing the structure of commutant
$Z(Q)$ and stabilizer~$I(Q)$ 
at any point $Q\in \Nilpp(d,\N)$, 
and their variation as families of superalgebras over~$\Nilpp(d,\N)$.

For instance, the space of first order deformations of $Q$---which is just the space
of solutions of $[Q,x]=0$ inside of $A^1$---is by definition the tangent
space to $\Nilp(d,\N)$, and equal to the odd part of the stabilizer:
\deq[eq:tangspace]{
T_Q\qty( \Nilp(d,\N) ) \cong I^1(Q) =  Z^1(Q),
}
where the last equality follows simply for degree reasons. 
Of course, $Q\in Z^1(Q)$ by nilpotence; we can therefore identify the (algebraic) 
tangent space to our projective variety at~$Q$ with the vector space $Z^1(Q)/Q$. It 
is important to note that, since the variety may be singular, the fiber dimension of 
the tangent bundle is not necessarily everywhere constant! It jumps up precisely along the singular locus.

To understand this better, we note that the identification \eqref{eq:tangspace} fits into 
the exact sequence
\deq[eq:es1]{
0 \longrightarrow Z^1(Q) \longrightarrow A^1 \overset{Q}{\longrightarrow} V 
\longrightarrow \coker(Q)^0 = V/E \longrightarrow 0.
}
In the context of twisted field theories, again, $E$ determines those vector fields on which
the correlators will not depend. Conversely, the cokernel $V/E$ should be thought of as 
the set of ``surviving'' momenta.
A central observation for the understanding of $Y(d,\N)$ is that $E$ is a 
{\it coistropic subspace} of $V$, i.e.\ $\langle v,w\rangle=0$ for all $v\in E$ 
implies that $w\in E$. This statement follows as a simple consequence of the nature 
of the pairing \eqref{eq:oddbracket} and will be explained below. Among other
things, it implies (again) that the minimal twists, for which $E$ is a half-dimensional
Lagrangian subspace of $V$, are the holomorphic ones.

A related sequence is the sequence defining the subalgebra 
$Z^0(Q)$. One can express this as follows:
\deq[eq:es2]{
0 \longrightarrow Z^0(Q) = \ker(Q) \longrightarrow A^0 \overset{Q}{\longrightarrow} A^1
\longrightarrow 
\coker(Q)^1 \longrightarrow 0.
} 
At face value, \eqref{eq:es2} is just the odd version of \eqref{eq:es1}.
However, the two sequences can also be merged together into a longer sequence
that is exact at every term except the middle one:
\deq[eq:longes]{
0 \longrightarrow Z^0(Q) \longrightarrow A^0 \overset{Q}{\longrightarrow} A^1
 \overset{Q}{\longrightarrow} V \longrightarrow V/E \longrightarrow 0.
}
The kernel of $Q$ in middle degree (i.e., in~$A^1$) is, as indicated before,  $Z^1(Q) \cong T_Q \widehat{Y}$. The image consists of those supercharges obtained by acting on~$Q$ with elements of~$A^0$; this is precisely the tangent space to the orbit of Lorentz and $R$-symmetry in which $Q$ sits, which is exactly the stratum in which it sits. {\it As a vector space,} we can also express the image of $Q$ in~$A^1$ as $A^0/Z^0(Q)$. The homology in middle degree is therefore the algebraic normal bundle to the pure stratum inside of the whole space~$\widehat{Y}$; the sequence is otherwise exact. 

One  therefore expects that jumps in dimension of~$V/E$ are connected to jumps in 
the dimension of~$Z^0(Q)$, and the two differ by an amount connected precisely to the 
codimension of the stratum.
Tautologically, the action of the unbroken bosonic symmetries (i.e., those in $Z^0$) 
commutes with $Q$. In particular, $E=\{Q,A^1\}$ carries a natural action of $Z^0$.
And the twisting map (or set of possible twisting maps) can be computed by just considering the appropriate subalgebras of~$Z^0(Q)_\R$.

 We emphasize that we study the sequence \eqref{eq:longes}, just like the varieties
$Y(d,\N)$, when $A$ is defined over the complex numbers. Parts of the sequence, however,
continue to make sense and are sometimes easier to understand over the reals.
For example, for a minimal twist in $d=2n$ dimensions, the real version of $Z^0(Q)$ 
contains a factor isomorphic to $\lie{u}(n)$, as appropriate in a holomorphic theory. In 
the complex version, $Z^0$ is not the complexification $\lie{gl}(n)$, but rather
a somewhat larger ``parabolic'' Lie algebra. This is related to a familiar
story in the context of homogeneous spaces; the reader will recall that
\deq{
P^1(\C) = SU(2)/U(1) = SL(2,\C)/B,
}
where $B$ is the subgroup of upper-triangular matrices rather than~$\C^\times$.

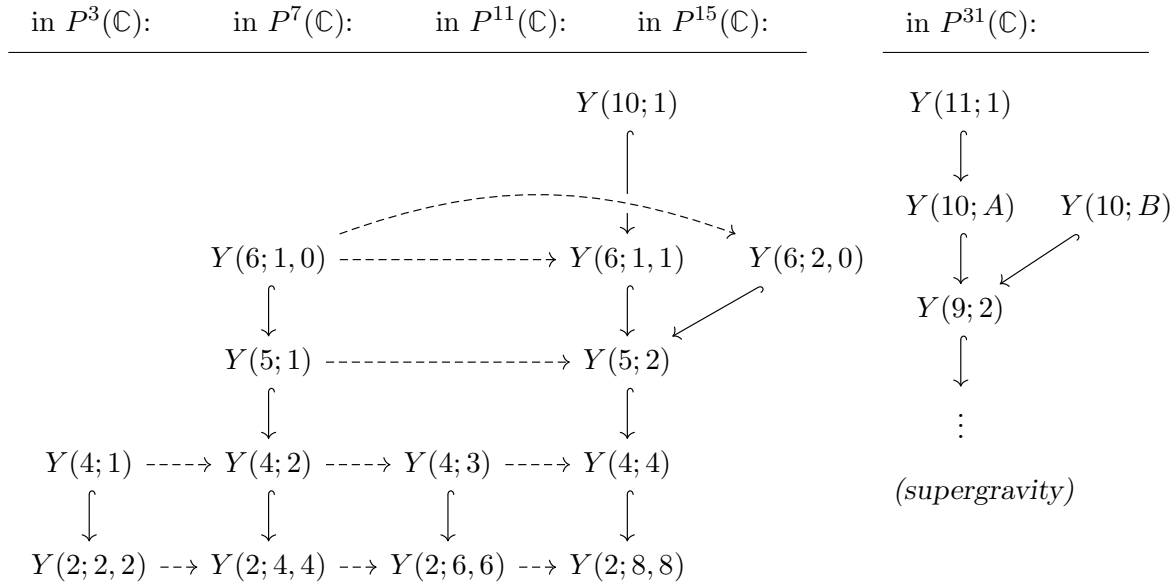
\begin{figure}
\begin{center}
\begin{tikzpicture}[mybox/.style={draw, inner sep=5pt}]
\node[anchor=west] at (0,0.6) (box){
\begin{tikzcd}[column sep = 0.5 cm]
\fixedwidth{\Nilpp(2,(2,2))}{\text{in $P^3(\C)$:}} & 
\fixedwidth{\Nilpp(2,(4,4))}{\text{in $P^7(\C)$:}} &
\fixedwidth{\Nilpp(2,(4,4))}{\text{in $P^{11}(\C)$:}}  & 
\fixedwidth{\Nilpp(2,(8,8)) }{\text{in $P^{15}(\C)$:}} 
\end{tikzcd}
};
\node[anchor=north west] at (0,0) (2box){
\begin{tikzcd}[column sep = 0.5 cm]
& & &   \Nilpp(10,1) \ar[dd,hook] & \\
& & &  & \\
& \Nilpp(6,1,0) \ar[d,hook] \arrow[crossing over, dashed, bend left=20]{rrr}{} \arrow[dashed]{rr}{}&  & \Nilpp(6,1,1) \ar[d,hook] &  \Nilpp(6,2,0) \ar[dl,hook] \\
& \Nilpp(5,1) \ar[d,hook] \arrow[dashed]{rr}{}& &\Nilpp(5,2) \ar[d,hook] & \\
\Nilpp(4,1) \ar[d,hook] \arrow[dashed]{r}{}&  \Nilpp(4,2) \ar[d,hook] \arrow[dashed]{r}{}&
\Nilpp(4,3) \arrow[dashed]{r}{} \ar[d,hook] &
 \Nilpp(4,4) \ar[d,hook] &   \\
\Nilpp(2,2,2)  \arrow[dashed]{r}{}&  \Nilpp(2,4,4) \arrow[dashed]{r}{}& \Nilpp(2,6,6) \arrow[dashed]{r}{}& \Nilpp(2,8,8) & 
\end{tikzcd} 
};

\node[anchor=west] at (11.5,0.6) (box){
\begin{tikzcd}[column sep = 0.5 cm]
\text{in $P^{31}(\C)$:}
\end{tikzcd}
};

\node[anchor=north west] at (11.5,0) (2box){
\begin{tikzcd}[column sep = 0.25 cm]
\Nilpp(11,1) \ar[d,hook] &  \\
\Nilpp(10,A) \ar[d,hook] & \Nilpp(10,B) \ar[dl,hook]  \\
\Nilpp(9,2) \ar[d,hook] &\\
\vdots & \\
\end{tikzcd} 
};

\node[anchor=west] at (11.5,-5.6) (box){\text{\sl (supergravity)}};
\draw (0,0.22) -- (10.5,0.22);
\draw (11.5,0.22) -- (15,0.22);
\end{tikzpicture}
\caption{Relations between nilpotence varieties}
\label{fig:nilpotencevarieties}
\end{center}
\end{figure}

Figure~\ref{fig:nilpotencevarieties} shows a diagram, listing relations between all of the varieties~$\Nilpp(d,\N)$ we will consider. (For compactness, we omit seven- through nine-dimensional supersymmetry algebras; the reader will find it obvious where these would sit in the diagram.) The diagram contains two types of inclusion arrows, vertical and horizontal. The vertical are straightforward to understand: upon dimensional reduction, one takes the higher-dimensional variety and throws out some of its defining quadrics. The number of supercharges (and hence the ambient projective space) does not change. Thus, the higher-dimensional variety is a subvariety of the one that appears upon dimensional reduction.

The horizontal are also reasonably clear, although of a slightly different nature. The key fact is that the set of nilpotent supercharges in an extended supersymmetry algebra, in which \emph{only} supercharges from a subalgebra with less supersymmetry appear, must be precisely the set of nilpotent supercharges in the smaller supersymmetry algebra of the same dimension. This means that the varieties with less-extended supersymmetry arise as hyperplane sections of the appropriate kind and dimension.

Considering the nilpotence varieties in different dimensions in unified fashion has several advantages.  For example, we will see that the five-dimensional nilpotence variety $\Nilpp(5,2)$ is the union of the ten-dimensional nilpotence variety $\Nilpp(10,1,0)$ and the six-dimensional nilpotence variety $\Nilpp(6,2).$  Similarly, we conjecture that the nine-dimensional nilpotence variety $\Nilpp(9,2)$ is the union of the eleven-dimensional nilpotence variety $\Nilpp(11,1)$ and the ten-dimensional nilpotence variety $\Nilpp(10,2,0)$.
These pairs of varieties are linked, or in ``liaison'' with each other, implying several subtle relations between them.
We can also understand twists (both topological and not) of theories with extended supersymmetry in lower dimensions as arising from dimensional reduction of holomorphic twists in higher-dimensional, minimally supersymmetric theories. See~\S\ref{ssec:dimred} below. 

In summary, nilpotence varieties are interesting and important objects for at least three reasons. Firstly, they are the natural moduli spaces of possible twists of Poincar\'e-invariant theories. Their strata enumerate these twists; they carry natural bundles encoding the breaking of symmetry in twisted theories; adjacency between the strata indicates when twisted theories admit further twists; and both dimensional reduction and extension of supersymmetry are manifested in natural relations between different nilpotence varieties. The bulk of the paper is dedicated to exploring these statements and studying concrete examples of nilpotence varieties in detail. 

Secondly, they allow one to consider ``global'' versions of twisting constructions; the pure-spinor superfield technique is one of these. It constructs representations of super-Poincar\'e algebras on spaces of fields, obtained by taking a canonical differential on the tensor product of two objects: one being a free superfield, considered as an $A$-module, and the other being any equivariant module over the coordinate ring $\OO \qty[ \Nilp(d,\N) ]$ of the affine nilpotence variety. 
This construction is intimately related to the Koszul complex over~$\OO \qty[ \Nilp(d,\N) ]$ of the cone point: the homology of that complex appears on the $E_1$ page of a spectral sequence which abuts to the BRST complex of a particular supermultiplet. We expand on these statements in~\S\ref{sec:PSSF}. Pursuing this program for the type IIB super-Poincar\'e algebra produces a BV complex with precisely the field content of type~IIB supergravity; see~\S\ref{ssec:IIB}.

Thirdly, they appear naturally just in the study of super-Poincar\'e algebras. The Chevalley--Eilenberg cohomology of the algebra of supertranslations can be shown to become a complex of $\OO \qty[ \Nilp(d,\N) ]$-modules, which in degree zero is just $\OO \qty[ \Nilp(d,\N) ]$; see~\S\ref{ssec:CEcoho}. 
We announce a computation of this cohomology for the $\N=(2,0)$ algebra in six dimensions; further results in this direction will appear in~\cite{twist-modules}.
Lie algebra cohomology of the full super-Poincar\'e algebra, studied among others by~\cite{Brandt,DixonMinasian,MovshevSchwarzXu}, can then be computed from this by an application of the Hochschild--Serre spectral sequence~\cite{HochschildSerre,MovshevHS}. As such, nilpotence varieties should be closely connected to recent work on ``brane scans,'' which (following older work~\cite{DAuriaFre}) computes the spectrum of extended objects sourcing form fields in supergravity theories from Chevalley--Eilenberg cohomology. See e.g.~\cite{DeAzcarragaTownsend,FiorenzaSatiSchreiber,HuertaSchreiber}. 

A couple of small remarks on the above: Firstly, in the context of supersymmetric gauge theory, one is \emph{a priori} interested, not in the entire theory of fields appearing in the Lagrangian, but in the truncation obtained by taking invariants of the (bosonic) gauge symmetry. To deal with this, though, one goes through the machinery of replacing this truncation by that to the invariants of a fermionic ``BRST'' symmetry.  Since one is already in this setting, one often speaks of ``adding $Q$ to the BRST differential.'' It is worth remarking that that procedure produces the $E_\infty$ page of the spectral sequence of the associated bicomplex, whereas taking the $Q$-invariants of the gauge invariants would produce the $E_2$ page. Of course, these don't necessarily agree; however, we are  aware neither of any concrete example in which they fail to, nor of a theorem that guarantees collapse at~$E_2$. (Such a theorem might be expected to use facts about the typical support of BRST or BV complexes in a limited range of ghost numbers, and therefore might fail for higher-spin gauge fields or other exotic theories.)

Secondly, we should emphasize that the role of the unbroken symmetry $I(Q)$ (or the corresponding twisting homomorphism) is indeed crucial. The full theory, with an action of the symmetry algebra~$A$, can of course only be formulated  on the flat spacetime~$\R^d$. (Translations, for example, will be broken in other situations.) So it does not make sense to say that a topological twist of the theory on some more general manifold~$M$ arises from taking invariants of the full theory on~$M$---the ``full theory on $M$'' does not admit an action of the algebra $A$, and this is therefore nonsensical. On the other hand, as is well-known, it is possible to formulate theories with some amount of unbroken supersymmetry on manifolds of special holonomy. We will take the perspective that the holonomy group must act through the algebra $I(Q)$ on the fields of the theory; as such, a theory for which $\lie{so}(d)' \subseteq I(Q)$ admits a formulation on any $d$-manifold, as one would expect for a topological field theory. By this reasoning, $2n$-dimensional twisted theories for which only $\lie{u}(n)\subset I(Q)$ cannot be defined on general topological $2n$-manifolds, but \emph{can} be expected to make sense on Calabi--Yau $n$-folds, and so on.

Lastly, we emphasize that the variety~$\Nilpp(d,\N)$ is the natural home for exactly those truncations of theories that take the form of $Q$-cohomology. While the vast majority of the invariants or restricted sectors of observables associated to supersymmetric theories are $Q$-cohomologies, some are not; the chiral ring of four-dimensional $\N=1$ theories, for example, proceeds by a seemingly slight variant of the same construction, which nonetheless is \emph{not} the $Q$-cohomology of any supercharge, and therefore is not related to a twist of the theory in any standard sense. (See~\cite{GNSSS} for further remarks.) It would be interesting to form constructions that generalize the chiral ring, to prove classification theorems for a family of such objects, or to find comparison results relating them to more standard $Q$-cohomologies. However, we do not take this path in the current work.

\paragraph*{Note:}
After this work was completed, \cite{Elliott} appeared while the manuscript was in preparation; a similar classification of twists in various dimensions is considered in that work, but with attention directed to the set of orbits or strata, rather than to the varieties themselves. A local formulation of twisting at the level of factorization algebras is also developed in that work. It would be interesting to understand spectral sequences between  various twisted theories rigorously in the context of that formalism.

\section{Preliminaries: Clifford modules and Cartan pure spinors}
\label{sec:spinors}

To develop a uniform notation with which to describe
examples explicitly, we digress to recall some standard facts about spinors, supersymmetry algebras, and the nature
of the decomposition in \eqref{eq:decomp}. For a more complete review, see~\cite{Deligne}, \cite{Georgi}, or the more computationally oriented discussion in~\cite{CY4}.

Let $V$ be the complexified defining representation of~$\lie{so}(d)$, with $\langle \cdot, \cdot \rangle$ its invariant nondegenerate bilinear form. For any $d$, the (complex) space of Dirac spinors of $V$ can be written
as
\deq[eq:dirac]{
S = \wedge^* L
}
where $L\subset V$ is a maximal isotropic subspace---that is, $\langle\cdot,
\cdot\rangle$ restricts to $0$ on $L$, and $L$ is of the greatest possible dimension
($n = \dim L = \lfloor d/2\rfloor$)
 with this property. 
This $2^n$-dimensional space $S$ carries 
a representation 
of the Clifford algebra $\Cl(V)$,
and in particular, a representation of $\lie{so}(V)\cong \wedge^2V$.

In the event that~$d$ is even, we have
\deq{
V=L\oplus L^\vee,
}
where $L^\vee$ is the dual to $L$ under the bilinear form.
$L$ acts on $S$ 
by wedging with a one-form and $L^\vee$ acts by contraction 
with a vector (i.e., by raising and lowering operators). 
For $d$ odd, we have
\deq{
V = L\oplus L^\vee\oplus \omega,
}
where $\omega=L^\perp/L$. Then it is clear that
\deq{
\wedge^*V= \wedge^*(L\oplus L^\vee) 
\oplus 
\wedge^{*}(L\oplus L^\vee)  \wedge\omega; 
}
 this is, as a vector space, identical with~$\Cl(V)$. $L\oplus L^\vee$ acts
on $S$ as in the even case, while $\omega$ is represented by $\pm(-1)^k$ on 
$\wedge^k L\subset S$.

As a representation of $\wedge^2V$, $S$ is irreducible when $d$ is odd, and
reducible when $d$ is even, in which case we write
\deq{S= S^+\oplus S^-}
for the two irreducible (chiral) pieces.
The representation respects the natural inner product on $S$ given by evaluation 
on the volume form,
\deq[eq:scalpair]{
(s,t):= \bigl(\text{degree-$n$ component of
$s\wedge t$}\bigr)}
which can therefore be viewed as a map of $\lie{so}(V)$ representations
$S\times S\to\C$.
When $d$ is even, and $d/2$ is odd, $(\cdot,\cdot)$ is a perfect
pairing between $S^+$ and $S^-$, while if $d=0\bmod 4$, it pairs
$S^\pm$ each with itself.
When $d$ is odd, no such distinction is possible, and $(\cdot,\cdot)$ is
a perfect pairing on $S$ by itself.

The map $S\times S\to V$ is described by in terms of $(\cdot,\cdot)$ by
duality:
\deq[eq:vecpair]{\langle\{s,t\},v\rangle := (s,vt)}
When $d=0\bmod 4$, the map restricts trivially to $S^+\times S^+$
and $S^-\times S^-$ and is non-trivial on $S^+\times S^-$. Vice-versa for
$d=2\bmod 4$. When $d$ is odd, there is again no distinction.

The description~\eqref{eq:dirac} furnishes an obvious solution to the
nilpotence equation $Q^2=0$: The one-dimensional subspace $\wedge^0L$ is
mapped by $V$ no further than $\wedge^0L\oplus \wedge^1L$. For $d>2$, this
implies $(Q,vQ)=0$ for all $v\in V$. 

However, to build our super-Poincar\'e algebra, we 
need a non-trivial pairing on $A^1$ as explained before. Thus, when
$d=0\bmod 4$, we need to include both $S^+$ and $S^-$.
Moreover, we have to respect the symmetry properties of
the pairings \eqref{eq:scalpair} and \eqref{eq:vecpair}. Their 
parity is given by
\deq{
(-1)^{{n(n-1)/2}}
}

When $d$ is even, we write
\deq[eq:evendec]{
A^1= \qty( S^+\otimes U^+ )  \oplus  \qty( S^-\otimes U^- ), 
}
with the existence of the symmetric pairing on $A^1$ restricting the dimensions 
$\N_\pm:=\dim U^\pm$ and the nature of the pairing on $U$ chosen according
to Table~\ref{tab:spinors}.

When $d$ is odd, we write
\deq[eq:odddec]{
A^1= S\otimes U,
}
with $\N:=\dim U$ and the pairing on $U$ again restricted according to Table~\ref{tab:spinors}.

\begin{table}[t]
\caption{Decomposition of~$A^1$ in Minkowski signature}
\begin{center}
\begin{tabular}{|c|c|c|c|c|c|}
\hline
$d$ mod 8 & pairing on $S$'s & pairing on~$U$'s  & notation & minimal~$A^1$ &  commutant\\
\hline
\hline
2 & $S^{+}$ and $S^{-}$ symmetric & orthogonal &  $\N_\pm$ independent  & $S^{+}$ & $\R$  \\
\hline
1, 3 & symmetric & orthogonal & $\N$ & $S$ & $\R$  \\
\hline
4, 0 & $S^{+}$ dual to $S^{-}$ & paired & $\N=\N_+=\N_-$ & $S^{+} \oplus S^{-}$ & $\C$  \\
\hline
5, 7 & antisymmetric & symplectic & $\N$ &  $S \otimes \C^2$ & $\bH$  \\
\hline
6 & $S^{+}$ and $S^{-}$ antisymmetric & symplectic &$\N_\pm$ independent & $S^{+} \otimes \C^2$ & $\bH$  \\
\hline
\end{tabular}
\end{center}
\label{tab:spinors}
\end{table}

\subsection{Stratification and rank}
\label{ssec:rankstrat}

As we pointed out above, $\Nilpp(d,\N)$ has an interesting stratification, which can be thought of as recording either different types of twist, jump loci for $Q$-cohomology, or the decomposition into orbits of Lorentz and $R$-symmetry; it will also be visible just in the structure of the variety itself, which will often be singular or reducible. In this section, we develop some tools which will help us to understand this stratification in later examples. 

For the purposes of the present work, we will be content with a rough intuitive notion of what a stratified space is; since all of the~$\Nilpp(d,\N)$ are concrete projective varieties, this will lead to no trouble. The reader with deeper interest in the topic is referred to~\cite{AFT} and references therein. Suffice it for now to remind the reader that a space is stratified when it is equipped with a continuous map to a poset $P$, which for us will always be finite. The poset is equipped with the \emph{poset topology}, in which a basis of open sets consists of the set 
\deq{
S_x = \{y:y> x\} \subseteq P.
}
(Equivalently, downward closures $\{y:y\leq x\}$ are closed.)
 A \emph{pure stratum} is the pullback over an element $p$ of the poset:
\begin{equation}
\begin{tikzcd}
Y_p \arrow[dashed]{r}{} \arrow{d}{} & Y \arrow{d}{} \\
p \arrow[hook]{r}{i} & P.
\end{tikzcd}
\end{equation}
Each pure stratum is required to be a smooth manifold of some fixed dimension; the dimension then defines a continuous map $\dim:P\rightarrow \mathbb{N}$ of partially ordered sets. Note that the continuity amounts to the statement that the composition of the two maps, as a map from~$Y$ to~$\mathbb{N}$, is lower semicontinuous. Note also that the cone on a $P$-stratified space is stratified over the cone on~$P$, obtained by adjoining a minimal element. 

Given the decomposition~\eqref{eq:decomp}, 
it is immediate to see that an element $Q\in A^1$ can always be thought of as either a matrix or a pair of matrices. Applying the anticommutator map uses the pairing on~$U$; one can therefore think of the result schematically as being an element in~$S \ttensor S$, where the nature and symmetry properties of the product depend on the dimension and amount of supersymmetry as above. Projection onto $V\subseteq S \ttensor S$, accomplished by the gamma matrices, should be understood. 

In dimensions four and six, this operation is closely related to usual matrix multiplication. In dimension four, for example, 
\deq{
A^1 = \qty( S^+ \otimes U ) \oplus \qty( S^- \otimes U^\vee),
}
where~$U=\C^\N$. The gamma matrices witness an isomorphism of representations $S^+ \otimes S^- \cong V$. Therefore, we can think of the nilpotence condition as stating that two matrices (a $2\times\N$ and an $\N\times 2$) multiply to zero. 

In six dimensions, on the other hand, 
\deq{
A^1 = \qty(S^+ \otimes U_+) \oplus \qty(S^- \otimes U_-),
}
and~$U$ is equipped with an \emph{antisymmetric} pairing $\omega$. The isomorphism is now between $V$ and~$\wedge^2 S^+$. This means that, for chiral supersymmetry---$\N=(1,0)$ or~$(2,0)$---one can think of the nilpotence condition as demanding that the product of matrices $Q\omega Q^t$ vanish identically. For $\N=(1,1)$ supersymmetry, the condition is instead that 
\deq[eq:6d11mat]{
Q_+ \omega Q_+^t + Q_- \omega Q_-^t = 0.
}
In both of these dimensions, and especially in dimension four, we will find it useful to think of the \emph{rank} of these matrices as an invariant determining the stratification of~$\Nilpp(d,\N)$. The poset over which nilpotence varieties in four dimensions are stratified is shown in Figure~\ref{fig:rankstrat} below. Of course, the rank of each matrix may not exceed two; it is also easy to see that the sum of the two ranks may not exceed~$\N$, since there would be no way to satisfy the nilpotence condition if this were true. The appropriate poset for each value of~$\N$ is the corresponding subposet of the one in the figure. The minimal element can be thought of as representing the cone point of the affine~$\Nilp(d,\N)$; it is absent for the projective variety.

The rank stratification in four dimensions is closely related to the zoo of harmonic superspaces \cite{Hartwell:1994rp}.  A textbook account is given in \cite{MR1865518}, based on the pioneering work of \cite{MR761034}.

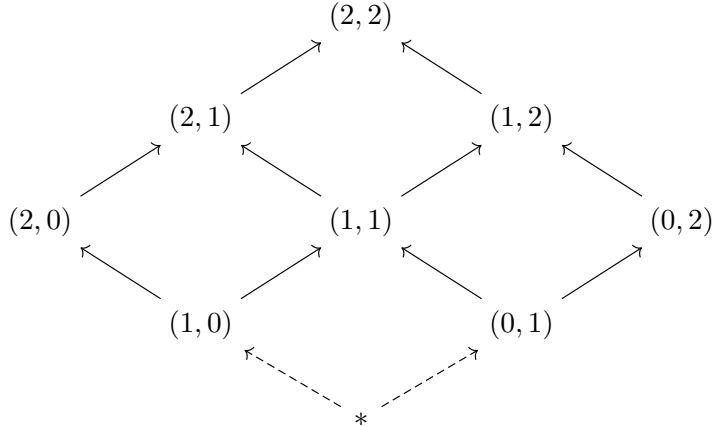
\begin{figure}[h]
\[
\begin{tikzcd}
  &   &    (2,2) & & \\
&  (2,1) \arrow[ru] &  & (1,2) \arrow[lu] & \\
          (2,0) \arrow[ru]  &  & (1,1) \arrow[ru] \arrow[lu]  & & (0,2) \arrow[ul]  \\
&  (1,0) \arrow[lu] \arrow[ru]  &  &  (0,1)  \arrow[lu] \arrow[ru]  & \\
  &   &    * \arrow[dashed,ru] \arrow[dashed,lu] & & \\
\end{tikzcd}
\]
\caption{Rank stratification in four-dimensional theories}
\label{fig:rankstrat}
\end{figure}

In other dimensions (eight and seven), we may also refer to the \emph{rank} of a given supercharge. The reader should always understand this to mean the rank of the matrix in~$S\ttensor U$ with which it is identified.

\subsection{Minimal supersymmetry and symmetric spaces}
\label{ssec:OG}

Using just the above generalities, we can understand the nature of~$\Nilpp(d,\N)$ (as well as~$Z(Q)$, the stabilizer algebra) for certain minimal supersymmetry algebras---namely, those where $d = \dim V$ is as large as it can possibly be given the dimension of~$A^1$. 

Recall that the Dirac spinor $S = \wedge^* L$ is a Clifford module, and is thus acted on by~$V$ as discussed above. In even dimensions, $S$ is reducible as a module of the \emph{even} Clifford algebra; $V$ then maps $S_+$ to~$S_-$, but the annihilator of a chiral spinor can still be defined. 

Given a nonzero element $Q$ of an irreducible spinor,
 its annihilator $\Ann(Q)$ is an isotropic subspace of $V$.  To show this, note that if~$v$ and~$w$ are in $\Ann(Q)$, then
\deq{
0 = \{v,w\}Q = 2 \langle v,w \rangle Q.
}
Hence $\langle v,w \rangle = 0$. By definition, $Q$ is a {\it Cartan pure spinor} when $\Ann(Q)$ is a maximal isotropic subspace of~$V$. 
The space of all Cartan pure spinors $\PS_0(V)$ is then a smooth projective variety in $P(S^\pm)$ (when $\dim(V)$ is even). In general, we could write
\deq{
\widehat{\PS}_k(V) = \{ Q \in S^+ : n - \dim \Ann(Q) \leq k \}.
}
This defines a family of projective varieties
\deq[eq:Cartanstrat]{
\PS_0 \subseteq \PS_1 \subseteq \dots \subseteq \PS_n = P(S^+ \cong \wedge^\text{ev}L),
}
where $\PS_k$ and~$\PS_{k+1}$ are not necessarily distinct. 

In dimensions four and six, all spinors are Cartan pure, and $\PS_0$ is just $P^1$ or~$P^3$ respectively.  In dimension eight, 
\deq{
\PS_0 = \lambda^6 \subset P^7,
}
where $\lambda^6$ is the zero locus of a single generic quadric in~$P^7.$ The stratification~\eqref{eq:Cartanstrat} up to dimension twelve is described by Igusa in~\cite{MR0277558}.

In fact, it is not hard to see that $\PS_0$ will always be related to the symmetric space
\deq{
\OG(n,d),
} 
which is the Grassmannian of isotropic $n$-planes in~$V$ with respect to its bilinear pairing. 
As above we will set $n=\lfloor d/2 \rfloor$, so that the maximal isotropic space $L^\vee = \Ann(Q) \subset V$ represents a point in~$\OG(n,d)$. 

In even dimensions, with real Euclidean signature, the choice of~$L$ corresponds to a choice of complex structure $\R^{2n} \cong \C^n$ on~$V_\R$. The unbroken Lorentz group is the subgroup $U(n) \subset SO(2n)$ leaving this complex structure invariant; the complexified version of this statement corresponds to a parabolic subalgebra of~$\lie{so}(2n,\C)$, preserving the flag $L\subset V$. But, by using the real statement, we can obtain the isomorphism
\deq{
\OG(n,2n) \cong SO(2n)/U(n),
}
where the objects at right are the usual (real!) Lie groups. In fact, one can also show 
 that, as varieties, $\OG(n-1,2n-1) \cong \OG(n,2n)$.

\eqref{eq:dirac} also makes clear a relation between the space~$\Nilpp(d,\N)$ and the orthogonal Grassmannian, in the case where~$A^1$ is just~$S^+$: 
Given any maximal isotropic subspace $L \subset V$, we can construct $S_+ = \wedge^\text{ev}(V)$; the element $Q = \wedge^0 V$ is, as we noted before, nilpotent. So there is a well-defined map
\deq{
\PS_0 \cong \OG(n,d) \rightarrow \Nilpp(d,\N).
}
An inverse to this map does not always exist. Given any nilpotent $Q$, we can reconstruct an isotropic subspace $\Ann(Q)$; there is, however, no guarantee that this subspace is maximal. We will see our first example of this in eleven dimensions. Note also that, by definition,
\deq{
\langle \{Q,s\},v \rangle = (s, vQ).
}
This makes it immediate that~$\Ann(Q) \subseteq E^\perp$. The opposite inclusion follows from nondegeneracy of the inner product on~$S$. Since $\Ann(Q)$ is isotropic, it follows immediately that~$E\subset V$ is coisotropic. (It is, of course, important to remember that $E$ and~$E^\perp$ need not be disjoint for complex vector spaces.)

Note that the assumption $A^1=S^+$ is satisfied for minimal supersymmetry only when $d=2\bmod 8$. When the existence of a symmetric bilinear pairing forces us to consider a non-trivial product $S\ttensor U$, the minimal nilpotence variety will be the product of $\OG(n,d)$ with some other space, related to the properties of $U$ in that dimension. 
We will make precise statements characterizing the nilpotence varieties of minimal supersymmetry algebras carefully below, in the dimensions (four, six, and ten) where they occur. 

\subsection{Dimensional reduction}
\label{ssec:dimred}

In many cases, we can think of extended supersymmetry algebras as having arisen by
dimensional reduction from higher dimensions, specifically from minimal supersymmetry in dimension four, six, or ten.
To describe the dimensional reduction procedure, we write $W$
for the fixed higher-dimensional spacetime (of dimension $2k$) and 
\deq{
W = V \oplus Z
}
for the orthogonal split into nondegenerate subspaces.
Under this reduction, we view $U$ in the decompositions \eqref{eq:evendec}
and \eqref{eq:odddec} as spinorial representations of $Z$ with its $\wedge^2 
Z$-invariant pairing determined with the same rules as before by the codimension 
$2k-d$. One checks, in the case $2k=10$, that there is a isomorphism of Lie algebras
$\lie{so}(Z)\cong R$ down to $d=4$; for smaller $d$, the physical $R$-symmetry may differ~\cite{Seiberg:1997ax}.

Dimensional reduction means setting the momentum generators in~$Z$ to zero; equivalently, one takes the quotient of the original algebra~$A$ by~$Z$. 
This doesn't affect $A^1$ or the pairing, so that $Q$ continues to define the same flag
\deq{
E^\perp = \Ann_W(Q)  \subset E \subset W.
}
But, to compute the space of exact translations in the dimensionally reduced theory, one has to take
\deq{
\Ann_V(Q) = \Ann_W(Q) \cap V;
}
this is obviously still an isotropic subspace of~$V$, but may not be maximal. Indeed, for minimal supersymmetry algebras, $\Ann_W(Q)$ is a $k$-dimensional Lagrangian of~$W$; the expected dimension of~$\Ann_V(Q)$ is therefore
\deq{
\dim \Ann_V(Q) = d - k,
}
which (since $d < 2k$) corresponds to a non-maximal isotropic. Of course, the subspaces~$V$ and~$\Ann_W(Q)$ may not intersect generically. The image of the dimensional reduction map will thus be subject to a finer stratification in the nilpotence variety of  the lower-dimensional algebra, including information about the dimension of the intersection.

When the intersection is generic, the commutant $Z^0(Q)_\R$ is also straightforward to compute. In the higher-dimensional supersymmetry algebra, the commutant is~$\lie{u}(k)$. The stabilizer of~$Z$ is~$\lie{so}(d)\times\lie{so}(2k-d)$. 
If $V_\R \subset W_\R$ is a complex subspace of dimension $n = d/2$ with respect to the complex structure induced by~$Q$ on~$W_\R$,
 the commutant is just $\lie{u}(n) \times \lie{u}(k-n)$. Generically, though, we will get a structure on~$V$ that looks like an identification
\deq{
V_\R \cong \C^{d-k} \times \R^{2k-d},
}
so that one ``topological'' direction is produced for each dimensionally reduced direction.
The corresponding unbroken subalgebra is
\deq{
Z^0(Q)_\R = \lie{u}(d-k) \times \lie{so}(2k-d) 
\subseteq \lie{u}(k) \subset \lie{so}(2k).
}
(Note that this formula holds when $d \geq k$.) 

One could also perform a similar computation, related to the realization of topological twists for brane worldvolume theories first proposed in~\cite{BershadskySadovVafa}. For flat branes in type~IIB 
string theory, these worldvolume theories are maximally supersymmetric Yang--Mills theories, and so can be constructed by the dimensional reduction procedure when $k=5$. The key fact is that, since~$A^1$ does not change under dimensional reduction, one could imagine a sort of reverse procedure, and consider the commutant of a nilpotent~$Q \in \Nilpp(d,\N_\text{max})$ inside of the ten-dimensional algebra~$A^0\cong\lie{so}(10)$---even though that $Q$ is not necessarily nilpotent in the ten-dimensional algebra itself. 

The key point of~\cite{BershadskySadovVafa} was that, since the $R$-symmetry space~$U$ of such a theory has a geometric interpretation as the fiber of the normal bundle to the brane, parallel transport also acts on $U$.
For a general cycle $C^d \subset X^{10}$ on which the brane is supported, the action of the Lorentz group will be twisted, so that the scalars of the theory (which, in flat space, parameterize normal displacements of the brane) transform correctly to be identified with the bundle describing a neighborhood of~$C$ in~$X$. 
We expect to be able to see the identification of twist maps with classes of string theory backgrounds in~\cite{BershadskySadovVafa} emerging just from the computation described above. However, we do not pursue the details of this argument in the present work. 

\section{Leitfaden: theories with four supercharges}
\label{sec:leitfaden}

In order to move smoothly into our discussion of examples, we'll begin with algebras that have four supercharges: $\N=1$ in four dimensions, and its dimensional reduction $\N=(2,2)$ in two dimensions. While these examples are simple enough to compute quickly by hand (and indeed have been written down before~\cite{GNSSS}, as well as probably being an implicit part of standard knowledge), we review them here in order to emphasize the key structural principles of the general story as simply and concretely as possible.
\subsection{Four dimensions: $\N=1$}
\label{sec:4dmin}

This is an example of a minimal supersymmetry algebra, which cannot arise by dimensional reduction; four is the maximal dimension in which an algebra with four supercharges can be defined. Therefore, based on the general argument given above in~\S\ref{ssec:OG}, one expects that
\deq{
\Nilpp(4,1) \sim  \OG(2,4) = SO(4)/U(2) = P^1(\C),
}
where the symbol $\sim$ indicates that $\Nilpp(4,1)$ may be the product of~$\OG(2,4)$ with an auxiliary space, as we indicated above. Here, since $d = 4 \pmod 8$, that auxiliary space is just two points, corresponding to the presence of two copies of the minimal spinor in the minimal supersymmetry algebra. Thus, we will find that in fact
\deq{
\Nilpp(4,1) = P^1(\C) \sqcup P^1(\C).
}
Indeed, it is not difficult to compute the variety explicitly and recover this. 
A general element of~$A^1$ is given by
\deq{
Q = u^\alpha Q_\alpha + v^{\dot\beta}\overline{Q}_{\dot\beta}
}
Here, just for variety, we use standard index notation, paralleling that in e.g.~\cite{WessBagger}. The supersymmetry algebra immediately gives rise to the four quadrics
\deq{
u^\alpha v^{\dot{\beta}} \Gamma_{\alpha\dot\beta}^\mu = 0,
}
which (to be perfectly explicit) are
\begin{align}
u^1 v^1 + u^2 v^2 &= 0, \nonumber \\
u^1 v^1 - u^2 v^2 &= 0, \nonumber \\
u^1 v^2 + u^2 v^1 &= 0, \nonumber \\
i\qty(  u^1 v^2 - u^2 v^1) &= 0.
\end{align}
Equivalently, one can write
\deq[4-1-eqs]{
u_1v_1 = u_2 v_2 = u_1 v_2 = u_2 v_2 = 0
}
(i.e., require each entry of the matrix $P_{\alpha\dot\beta}$ to vanish.)
These imply either $u=0$ or $v=0$, so we get two nonintersecting copies of $P^1(\C)$ in $P^3(\C)$.

This could have been seen easily from our discussion in~\S\ref{ssec:rankstrat}. Here, the statement is that we have a two-by-one and a one-by-two matrix that multiply to zero. This immediately implies that one of the two matrices must vanish identically, reproducing the result computed above.

Now, each $P^1(\C)$ component looks like a family of complex structures on~$\R^4$; the two components correspond to complex structures inducing different orientations on~$\R^4$. The tangent space to a given component is also simple to understand: it just means that a chiral supercharge may be deformed by the \emph{other} supercharge of the same chirality, while preserving nilpotence. (Of course, this deformation is trivial up to a Lorentz transformation.)

The case of three dimensions is essentially an afterthought. For the general reasons discussed in~\S\ref{ssec:OG}, the variety is 
 exactly the same: 
two disjoint copies of $P^1$, which is just two copies of $\OG(1,3)\cong \OG(2,4)$. 
(At the level of schemes, though, there is one important subtlety: the dimensionally reduced variety is defined by one fewer quadric, so that the Hilbert series will differ precisely by subtracting $t^2$. This may have some effect on the pure spinor superfield formalism; we discuss this further below.)

\subsection{Two dimensions: $\N=(2,2)$}

The simplest way to obtain the relevant set of equations is by dimensional reduction.
One throws out two of the above quadrics, obtaining
\deq{
u_1v_1 = u_2 v_2 = 0.
}
And a moment's thought shows that this consists of four lines in $P^3(\C)$, intersecting pairwise. Two such correspond to $A$-twists, and two such to $B$-twists; they intersect in points (corresponding to a single chiral supercharge that is also an $R$-symmetry eigenstate) that define the holomorphic twists related to the elliptic genus in two dimensions. One can recall that such theories in principle have four different chiral rings~\cite{LercheVafaWarner}, originally called $(c,c)$, $(c,a)$, $(a,c)$, and~$(a,a)$; however, they are pairwise equivalent and define only two distinct truncations.

Several new phenomena appear here for the first time: first off, the variety has several irreducible components (or, more generally, strata), corresponding to different classes of twists. Secondly, it has a singular locus, at which the holomorphic (minimal) twists appear. Note that the holomorphic locus consists simply of points; one should remember that it now corresponds to the space of complex structures on~$\R^2$, which is just a point,
\deq{
SO(2)/U(1) \cong *.
}
However, the lowest stratum consists of four such points, since $\N=(2,2)$ supersymmetry is four times as large as the minimal supersymmetry algebra ($\N=(0,1)$) in two dimensions.

In fact, the singular nature of the tangent space at a holomorphic point is easy to see from a computation in the supersymmetry algebra. One may deform a left-moving supercharge by either of the right-moving $R$-symmetry eigenstates, but \emph{not} by any linear combination thereof.

Additionally, two of the four components of our variety arise by dimensional reduction, as the image of the natural map
\deq[eq:dimredB]{
P^1(\C) \sqcup P^1(\C) = \Nilpp(4,1) \hookrightarrow \Nilpp(2,2,2).
}
Quite abstractly, along the lines of the general discussion in~\S\ref{ssec:dimred}, one expects (upon dimensionally reducing the holomorphic twist in four dimensions) to produce, for generic points in the stratum, a purely topological twist in two dimensions. And, indeed, this is true: the two projective lines corresponding to the $B$-twist are the image of the map~\eqref{eq:dimredB}. The new components, appearing for the first time in two dimensions, correspond to the $A$-model topological twist.

We will see similar phenomena (stratification, reducibility, singularities) occurring in more complicated ways in higher dimensions, but all of the essential features are already present here. Other nilpotence varieties in dimension two are simple to describe: The nilpotence constraint leads to a single quadric in the left-moving set of variables, and a single quadric in the right-moving set of variables. The defining ideal consists of this pair of quadrics; the singular locus is the intersection of the variety with the chiral coordinate planes, and consists of the disjoint union of a single quadric in each space. 

\section{Minimal supersymmetry: dimensions ten and six}

\subsection{Ten-dimensional minimal supersymmetry: Berkovits' pure spinor space}

In standard coordinate notation, the ten-dimensional supersymmetry algebra is 
\deq{
\{ Q_\alpha, Q_\beta \} = \gamma^I_{\alpha\beta} P_I,
}
where $Q_\alpha$ is a Weyl spinor in the $S_-$ representation. In our previous notation, $U_- = \C$ and~$U_+ = 0$. 
Using this algebra, one trivially computes that a supercharge $Q$ is nilpotent precisely when
\deq[eq:PSconstraint]{
\{ Q, Q\} = \left( u^\alpha \gamma^I_{\alpha\beta} u^\beta \right) P_I = 0.
}
Based on previous discussion, it is straightforward to see that the space of solutions to the pure-spinor constraint is a ten-dimensional projective variety,
\deq{
\Nilpp(10,1) \cong SO(10)/U(5) \subset P^{15}(\C).
}
(By abuse of notation, $(10;1)$ here refers to $(10;1,0)$.) The stratification is trivial, in this case: the algebra admits only minimal (holomorphic) twists. 

The next question to ask is, what does the tangent space to the space of pure spinors at a point look like, inside of the space of all supercharges, in terms of the decomposition of spinors induced by the reduction of the Lorentz group corresponding to the chosen point? In other words, what are infinitesimal parameters $u^\alpha$ such that $Q + u^\alpha Q_\alpha$ is still nilpotent, at linear order in the $u$'s?

As we've seen repeatedly, choosing a nilpotent $Q$ fixes a reduction of the real Lorentz group from $SO(10)$ to $U(5)$, corresponding to a choice of complex structure. The spinor representation branches as~$\wedge^\text{ev}{L}$, where~$L$ carries the fundamental representation of~$U(5)\cong L \otimes L^\vee$. In other words,
\deq{
S_- =   \mathbf{1}^{-5/2} + \rep{10}^{-1/2} +  \crep{5}^{3/2}.
}
We'll adopt a standard index notation ($Q$, ${Q}^{mn}$, and~$Q_m$) for the decomposition of the $S_-$ spinor, and $P^m$, ${P}_{m}$ for that of the vector; here, raised and lowered indices correspond to fundamental and antifundamental tensor indices for~$SU(5)$. Working out what the commutation relations of the algebra reduce to, one finds
\deq{
\{Q, Q\} = 0, \quad \{Q, Q_j \} = P_j,
\quad
\{Q,{Q}^{jk} \} = 0,
}
\deq{
 \{Q_j, Q_k \} = 0, \quad
  \{ Q_j, {Q}^{kl} \} \sim \delta_j^k P^{l} - \delta_j^l P^k, \quad
  \{ {Q}^{j k}, {Q}^{l m} \} \sim \epsilon^{jklmn} P_n.
}
From the first line, it follows immediately that 
\deq{
Z^1(Q) \cong T_Q \Nilp(10,1) \cong \rep{1} \oplus \rep{10}.
}
Furthermore, going to quadratic order in the algebra, one can see that a supercharge $Q + u^m Q_m + u_{mn} Q^{mn}$ is nilpotent if and only if
\deq[eq:PSsolv]{
u^m + \epsilon^{mnpqr}u_{np}u_{qr} = 0.
}
This gives explicit coordinates on a neighborhood of~$Q\subset\Nilpp(10,1)$. 
Another equation arises from the holomorphic momentum, but it is implied by~\eqref{eq:PSsolv}~\cite{Cederwall:2011yp}.  The ten defining equations of $OG(5,10)_{+}$ are given explicitly in \cite{MR1363081}.

The nilpotence variety $\Nilpp(9,1)$ is, of course, identical, up to the small subtlety mentioned before. We will return to dimensional reductions to dimensions eight and seven below in~\S\ref{sec:dimred}.

\subsection{Six-dimensional minimal supersymmetry}

In the six-dimensional minimal supersymmetry algebra, $(\N_+,\N_-) = (1,0)$; the supercharges sit in the representation $S^+ \otimes U_+$ with~$U_+ = \C^2$. (This is the smallest possible $U$ admitting a symplectic pairing; it is thus not possible to construct an algebra in six dimensions with only four supercharges.) 

A basis can thus can be written in the form~$Q_\alpha^i$, where $\alpha$ is an $S^+$ spinor index of $SO(6)$ (the $\rep{4}$ of $SU(4)$), and $i$ is an index for the $\rep{2}$ of~$SU(2)$.
In such coordinates, the algebra is:
\deq{
\{ Q^i_\alpha, Q^j_\beta \} = \varepsilon^{ij} \Gamma^\mu_{\alpha\beta} P_\mu.
}
From this, one can derive the collection of six quadratic equations
\deq{
u^{\alpha}_{i} u^{\beta}_{ j} \Gamma^\mu_{\alpha\beta} \epsilon^{ij} = 0,
}
which (as indicated above in~\S\ref{ssec:rankstrat}) can be conveniently represented as the $2\times 2$ minors of the following matrix:
\deq{
\begin{bmatrix}
u^1_1 & u^2_1 & u^3_1 & u^4_1 \\
u^1_2 & u^2_2 & u^3_2 & u^4_2 
\end{bmatrix} = \begin{bmatrix} u^\alpha_1 \\ u^\alpha_2 \end{bmatrix}.
}
So~$\Nilpp(6,1,0)$ is a determinantal variety: the space of rank-one matrices in $M^{2\times 4}(\C)$.
Alternatively, one could say that it is the Segre embedding $\sigma(P^1\times P^3)$ 
of $P^1 \times P^3$ in $P^7$. One way to see this is to recall that the data of a rank-one matrix determines both a line in~$\C^2$ (its kernel) and a line in~$\C^4$ (its image). 

As one would expect from~\S\ref{ssec:OG}, $P^3 = SO(6)/U(3) = \OG(3,6)$ appears as a factor, and the space is smooth. The additional $P^1$ factor is due to the fact that, as we noted above, no supersymmetry algebra can be constructed with a single copy of the minimal spinor of~$SO(6)$. So there is an ``unavoidable'' $SU(2)$ $R$-symmetry, even though no extended supersymmetry is present. The $P^1$ is just the projectivization of~$U$. 

This nilpotence variety can be used in the pure-spinor superfield technique, as in~\S\ref{sec:PSSF}, to produce the vector multiplet in six-dimensional super Yang--Mills theory from the cohomology of the structure sheaf. This computation was performed in~\cite{Cederwall-6d}.

The five-dimensional $\N=1$ variety $\Nilp(5,1)$ is 
$\OG(2,5) \times P^1 \cong P^3 \times P^1$,
which is identical to the six-dimensional nilpotence variety $\Nilpp(6,1,0)$. 
The saturation of its defining ideal is the defining ideal in six dimensions.
The supersymmetric partition function of the twisted theory on contact manifolds was computed in~\cite{Kallen:2012cs} using supersymmetric localization, inspired by previous work~\cite{Baulieu:1997nj}.

Upon reducing to four dimensions, though, one gets the $\N=2$ variety, which---just as we found upon reducing four-dimensional minimal supersymmetry to two dimensions---is in fact different. We turn now to analyze this example.

\section{Extended supersymmetry in four and fewer dimensions}

\subsection{Four-dimensional $\N=2$}
For the first time, we encounter a genuinely interesting reducible variety. It can be obtained either by throwing out two quadrics from the above six, or by looking directly at the four-dimensional $\N=2$ algebra, perhaps studying it in terms of matrices as discussed in~\S\ref{sec:4dmin}. We discuss it from each of these perspectives in turn.

Upon dimensional reduction, $R$-symmetry stays $R$-symmetry; the spinor of $SO(6)$ decomposes as the Dirac spinor $(\rep{2},1)\oplus (1,\rep{2})$ of~$SO(4)$.
In standard coordinate notation, we could write $A^1$ as
\deq{
\begin{bmatrix}
u_1^1 & u_1^2 & {v}^{\dot 1}_1 & {v}^{\dot 2}_1 \\
u_2^1 & u_2^2 & {v}^{\dot 2}_2 & {v}^{\dot 2}_2 
\end{bmatrix} = \begin{bmatrix} u^\alpha_1 \\ u^\alpha_2 \end{bmatrix}.
}
We've thrown out two of the six minors of this matrix, though: the relevant minors for the dimensional reduction all mix left- and right-chiral spinors, and there are four such. This means that new components of the variety can arise: they are precisely $u=0$ and $v=0$. Off of these loci, the matrix must have rank one, and we just recover the Segre variety arising from dimensional reduction.

The variety thus has three irreducible components. Two are just nonintersecting coordinate $P^3$'s in $P^7$, and the third is~$\Nilpp(6,1)$.
The intersections of the coordinate $P^3$'s with the third component are along $P^1 \times P^1$'s.
Again, based on general ideas about dimensional reduction of holomorphic twists, we expect the image of~$\Nilpp(6,1)$ to include twists that are holomorphic in one and topological in two directions. And this is indeed the case.

These components can be fruitfully be organized by thinking about the stratification of the variety by rank, when formulating the choice of supercharge in terms of a pair of matrices as discussed above. Since $\N=2$, we are interested in a pair of two-by-two matrices that multiply to zero; the sum of the two ranks cannot exceed two, so that the rank stratification poset looks as follows:
\begin{equation}
\begin{tikzcd}[row sep = 1 em, column sep = 1 em]
          (2,0) 
          &  & (1,1) 
          & & (0,2) 
           \\
&  (1,0) \arrow[lu] \arrow[ru]  &  &  (0,1)  \arrow[lu] \arrow[ru]  & \\
  &   &    * \arrow[dashed,ru] \arrow[dashed,lu] & & \\
\end{tikzcd}
\end{equation}
Here the $*$ can again be thought of as representing the (absent) cone point. Recall, now, that each maximal element corresponds to an irreducible component of the variety, which is its downward closure. The $(2,0)$ components, as affine spaces, are obviously $\C^4 \subset \C^8$; its projectivization is thus $P^3$, and the $(1,0)$ component, as a projective variety, is $P^1\times P^1$---one factor corresponding to the image of the map and one factor to its kernel. (This can also be seen by noting that the determinant is a generic homogeneous quadric in four variables, and such a quadric in $P^3$ cuts out a $P^1\times P^1$.)

The stratification thus contains the following pieces:
\begin{equation}
\begin{tikzcd}[row sep = 2 em, column sep = -1 em]
          \text{left DW} 
                    &  & \text{Kapustin half-holo.} 
          & & \text{right DW} 
           \\
& \text{left holo.} \arrow[lu] \arrow[ru]  &  & \text{right holo.}  \arrow[lu] \arrow[ru]  & 
\end{tikzcd}
\end{equation}
A similar pattern will recur in eight dimensions: we will obtain the ten-dimensional nilpotence variety, together with two disjoint coordinate planes. 

\subsection{Dimensional reduction to $\N=4$ in three dimensions}

The Lorentz group is $SO(3)$, which we think of (after passing to the spin group) as~$SU(2)$. 
The $R$-symmetry is $SO(4)\cong SU(2)\times SU(2)$; one factor can be thought of as coming from the $R$-symmetry in six dimensions, and another from $SO(3)$ rotations in the three transverse coordinates. A peculiarity of three-dimensional $\N=4$ theories is that the two $SU(2)$ factors in the $R$-symmetry group, although they appear symmetrically in the algebra, do \emph{not} appear on an equal footing with regard to the structure of standard multiplets. 

With respect to dimensional reduction from four dimensions, the Lorentz group $SO(3)$ is the diagonal subgroup inside of the four-dimensional $SU(2)\times SU(2) \cong SO(4)$ Lorentz symmetry.
So the relevant group action is $SU(2)^3$ in both four and three dimensions---but the group does not act in the same way. 

The nilpotence variety $\Nilpp(3,4)$ consists of two copies of Segre-embedded $P^1 \times P^3$. One of these is the copy of~$\Nilpp(6,1)$ coming from dimensional reduction; the other contains both coordinate $P^3$'s of~$\Nilpp(4,2)$. The intersection of the two components (which is the singular locus of the variety) is Segre-embedded $P^1 \times P^1 \times P^1$.  This suggests organizing the eight parameters in a three-index tensor, corresponding to the weight labeling for the representation $(\rep{2},\rep{2},\rep{2})$ of~$\lie{su}(2)^{\oplus 3}\subset A^0$.

A  very similar structure will recur in the seven-dimensional nilpotence variety. There, dimensional reduction from ten dimensions produces one component; the other is a Segre-embedded product of projective spaces, containing both of the coordinate planes from dimension eight.

\subsection{Four-dimensional $\N=3$ supersymmetry}
In matrix language, we now have a pair of $2\times 3$ matrices, which are required to multiply (along the $R$-symmetry indices) to zero. The constraint on the rank now implies that $r_+ + r_- \leq 3$, so that the rank poset is truncated. The appropriate truncation looks as follows:
\begin{equation}
\begin{tikzcd}[row sep = 1 em, column sep = 1 em]
&  (2,1)
 &  & (1,2) 
 & \\
          (2,0) \arrow[ru]  
          &  & (1,1) \arrow[ru] \arrow[lu]  
          & & (0,2) \arrow[ul] 
           \\
&  (1,0) \arrow[lu] \arrow[ru]  &  &  (0,1)  \arrow[lu] \arrow[ru]  & \\
\end{tikzcd}
\end{equation}
Based on this, together with the remarks about irreducible components made above, we can immediately see that one expects the variety to have two (isomorphic) irreducible components. We can schematically indicate the form of the stratification as follows:
\begin{equation}
\begin{tikzcd}[row sep = 1 em, column sep = 1 em]
&  B_+^7  &  & B_-^7  & \\
          P^5 \arrow[ru]   &  & E^6 \arrow[lu]  \arrow[ru]  & & P^5 \arrow[lu]  \\
&  P^1 \times P^2    \arrow[ru] \ar[lu] &  & P^1 \times P^2 \arrow[lu] \ar[ru] & \\
\end{tikzcd}
\end{equation}
Here, entries in the table represent the downward closures $Y_{\leq p}$ of each stratum, rather than the pure strata $Y_p$ themselves; thus, the arrows represent the inclusion maps $Y_{\leq p} \hookrightarrow Y_{\leq p'}$ for any poset elements $p'>p$.

The nilpotence variety has two irreducible components $B_\pm$.  They are both given by the intersection of 7 (nongeneric) quadrics in $P^{11}.$
Their intersection is $E = B_+ \cap B_-$.  This space looks like the intersection of~$\OG(5,10)$ with four (generic) hyperplanes.  The ten equations come from the four defining equations and two sets of three equations for the vanishing of the minors.  
The Hilbert series of~$E$ is
\deq{
\frac{(1 + 4t + t^2)}{(1-t)^4}\frac{(1-t^2)}{(1-t)^4}.
} 
We can obtain a description of the pure strata by using canonical factorizations of the supercharges, thought of as linear maps. For example, suppose that the rank is $(2,1)$. Then we can write the factorization
\begin{equation}
\begin{tikzcd}
&U / \ker Q_- \arrow{r}{GL(1)} & \im Q_- \arrow[hookrightarrow]{d}{} \\
S_+ \arrow{r}{Q_+} \arrow[twoheadrightarrow]{d}{=} & U\arrow[twoheadrightarrow]{u}{} \arrow{r}{Q_-} & S_- .\\
S_+/\ker Q_+ \arrow{r}{GL(2)} & \im Q_+ = \ker Q_- \arrow[hookrightarrow]{u}{} & 
\end{tikzcd}
\end{equation}
From this, there is an apparent affine description of the pure stratum:
\deq{
GL(2) \times  GL(1) \hookrightarrow 
\widehat{Y}_{(2,1)} \twoheadrightarrow
P^2 \times P^1.
}
The factors that are projective spaces (more generally, they will be Grassmannians or flag varieties) come from the choices of kernel and image in the three columns of the diagram, whereas the  $GL$ groups come from choices of isomorphism along the horizontal maps of the diagram. 

After taking the quotient by rescaling, which acts simultaneously on the $GL$ factors, one can set the $GL(1)$ factor to unity, obtaining the description
\deq{
GL(2) \hookrightarrow
{Y}_{(2,1)} \twoheadrightarrow
  P^2 \times P^1
}
of the open stratum $B^7 \setminus (E^6 \cup P^5)$. It is straightforward to see that the dimension agrees. 

For the case of rank $(1,1)$, we would write a very similar diagram:
\begin{equation}
\begin{tikzcd}
&U / \ker Q_- \arrow{r}{GL(1)} & \im Q_- \arrow[hookrightarrow]{d}{} \\
S_+ \arrow{r}{Q_+} \arrow[twoheadrightarrow]{d}{} & U\arrow[twoheadrightarrow]{u}{} \arrow{r}{Q_-} & S_- .\\
S_+/\ker Q_+ \arrow{r}{GL(1)} & \im Q_+  \arrow[hookrightarrow]{u}{} & 
\end{tikzcd}
\end{equation}
Examining the columns and rows as above, we would compute that 
\begin{equation}
\begin{tikzcd}
P\qty[GL(1)\times GL(1)] \arrow[hookrightarrow]{r}{}
& E^6 = {Y}_{\leq (1,1)} \supseteq {Y}_{(1,1)}  \arrow[twoheadrightarrow]{d}{}\\
& P^1 \times \Fl(1,2;\C^3) \times P^1.
\end{tikzcd}
\end{equation}
Now, the fiber (after dividing out by projective equivalence) is just $\C^\times$. But it is more properly thought of as a copy of~$P^1(\C)$, where the poles correspond to the rank dropping from $(1,1)$ to $(1,0)$ or~$(0,1)$. The variety $E^6$ can be thought of as a family over this~$P^1$, with generic fiber $P^1 \times P^1 \times \Fl(1,2;\C^3)$. At the poles, however, the fiber drops to $P^2 \times P^1$; the map is one of the two forgetful maps from the full flag variety to~$P^2$, together with projection from the corresponding $P^1$ to a point.

The factor $\Fl(1,2;\C^3)$ is part of $\N = 3$ harmonic superspace.  It is also related to Witten's twistor interpretation of classical $\N=3$ super Yang--Mills theory \cite{Witten:1978xx, Movshev:2004ub}.

Much of the interpretation of the strata is analogous to that of the case of $\N=2$ supersymmetry. For example, the strata $(1,0)$ and~$(0,1)$ still correspond to the holomorphic twists, and the $(1,1)$ stratum to a half-holomorphic, half-topological twist. The $P^5$ strata are twists of Donaldson--Witten type, also obtained by regarding the $\N=3$ theory as an~$\N=2$ theory. The only conceptually new ingredient is the $(2,1)$ stratum, which appears to correspond to a weakly topological twist of four-dimensional $\N=3$ theories. As far as we know, this twist has not been considered in the literature before. Of course, it is worth mentioning that every $CPT$-invariant $\N=3$ theory of matter and gauge fields is in fact an $\N=4$ theory. Thus, the only interesting theories that are strictly $\N=3$ supersymmetric are supergravity theories.

We will return to the case of maximal supersymmetry in four dimensions after examining other theories with sixteen supercharges.

\section{Dimensional reduction: dimensions nine, eight and seven}
\label{sec:dimred}

We now consider dimensional reductions from the pure spinor space $\OG(5,10)$ in ten dimensions. As mentioned before, the story will bear some striking parallels to dimensional reductions of six-dimensional minimal supersymmetry. 

In nine dimensions, as before, there is not much to say. 
The saturation of the defining ideal for the nilpotence variety in nine dimensions is equal to the defining ideal of the ten-dimensional nilpotence variety.  There is a single type of twist, which is holomorphic in eight dimensions and topological in one. This is captured by the classical isomorphism
\deq{
\OG(4,9) \cong \OG(5,10)
}
at the level of varieties. It may be, though, that the algebraic distinction between the two cases has implications for the pure spinor superfield formalism.

Recall that the 10-dimensional nilpotence variety $OG(5,10)$ is described by 10 quadrics.  Viewing $V$ as $\bf{5} \oplus \overline{\bf{5}}$ we call these quadrics $Q_i$ and $Q^i.$  
The 9-dimensional nilpotence variety $\Nilpp(9,1)$ is defined by the ideal generated by $Q_{i}, Q^{i}$ $i=0,1,\dots, 4$ and a linear combination of $Q_{5}$ and $Q^5.$  The quadrics satisfy the relation
$$Q_iQ^i = 0.$$  This implies that $Q_5 Q^5 = 0$ and hence the square of the other linear combination of $Q_{5}$ and $Q^5$ must vanish.

\subsection{Eight dimensions, $\Spin(7)$, and division algebras}

Analogous to $\Nilpp(4,2)$, we obtain a variety with three irreducible components. One of these is just $\Nilpp(10,1)$, appearing upon dimensional reduction; the others are disjoint coordinate $P^7$'s. Each of these $P^7$ components meets $\Nilpp(10,1)$ along the space
\deq{
\OG(4,8) \cong \qu(6) \subset P^7. 
}
Here, $\qu(6)$ means the zero set of a single generic quadric in~$P^7$; this is the space $SO(8)/U(4)$ of holomorphic twists, as we would have expected. 
The open strata of the $P_{\pm}^7$ components correspond to an 8d $\Spin(7)$ twisted theory~\cite{BaulieuKannoSinger}. The open stratum of other component, $\OG(5,10)$, corresponds (by the general logic given in~\S\ref{ssec:dimred}) to a twist which is holomorphic in three and topological in two dimensions. 

For the two $P^7$ components, the commutant is easy to compute. The spinor $S$ is sixteen dimensional; it decomposes as the sum of the Weyl spinors $S^+ \oplus S^-$ of~$SO(8)$. We have $U_+ = (U_-)^\vee = \C$; just as in four dimensions, the pairing is between~$S_+$ and~$S_-$. This implies immediately that any element lying in~$S_\pm\subset A^1\cong S$ is nilpotent; these are the two $P^7$ components we identified above. Now, the stabilizer of an element of~$S^+ \otimes U_+$ is   just the stabilizer of an element of~$S^+$ in~$SO(8)$, together with the stabilizer of an element of~$U_+$ in~$R$. The former, by definition, is the exceptional embedding of~$\Spin(7)$ in~$SO(8)$; the latter is trivial, since  the $R$-symmetry is just $U(1)$ in this case. The exceptional $\Spin(7)$ embeddings can be thought of as arising from the standard embedding after a triality transformation that exchanges the roles of the vector with one of the Weyl spinors. Since there are two Weyl spinors, there are two exceptional $\Spin(7)$ embeddings, corresponding to our two $P^7$ components.
In this case, $E=V$, since both the vector and one of the Weyl spinors of~$SO(8)$ branch to the irreducible Dirac spinor of~$\Spin(7)$; the image of this spinor under $\ad_Q$ is then all of~$V$. We are thus in the case of a weakly topological twist.

For the~$\OG(5,10)$ component, which mixes spinors of different chirality, the commutant is most easily computed by dimensional reduction. In the real Lorentz algebra, it is~$U(3)\times SO(2) \subseteq SO(8)$.


It is also worth remarking that we can give a uniform description of nilpotence varieties in dimensions 3, 4, 6, and~10 in terms of the real division algebras. As is well-known~\cite{Deligne,CederwallDivAlg,BaezDivAlg}, if~$D$ is a normed division algebra over~$\R$, the spin groups $\Spin(\dim D + 2)$ correspond to the groups $SL(2,D)$, and the spinor representation can be thought of as corresponding to~$D^2$. In fact, we can formulate the equations of the corresponding nilpotence variety in terms of the division algebra. If $(a,b) \in D^2$, then the equations take the form
\deq{
a \bar{a}  = b \bar{b} = a \bar{ b} = 0.
}
Since the first two are real, this is a total of $\dim D + 2$ quadrics. Additionally, one should interpret the equations inside of $D \otimes_\R \C$---if one did not complexify, the solution set would be vacuous. Of course, the three-dimensional $\N=1$ variety is just the empty set; the symmetric square of the spinor of~$SO(3)$ is precisely the vector representation, so that no nilpotent supercharges can be found.

Upon dimensional reduction by two (i.e., to $\dim D$ dimensions), we can discard the two constraints on the norm, preserving the single equation
\deq{
a \bar{ b} = 0.
}
It is easy to check that this generates two new irreducible components, the coordinate planes $a=0$ and $b=0$, as we have seen explicitly in two, four, and eight dimensions upon reduction from four, six, and ten respectively.

\subsection{Seven dimensions and $G_2$}

The spinor $S$ of~$SO(7)$ is eight-dimensional and irreducible. The $R$-symmetry space $U$ is~$\C^2$, equipped with a symplectic pairing. (This is therefore the minimal supersymmetry algebra in dimension seven.) 

Thinking of  $Q \in S \otimes U$ as an $8 \times 2$ matrix, there is an immediate way to satisfy the nilpotence condition: simply take $Q$ to have rank one. Then $Q$ is the tensor product of an element of~$S$ with an element of~$U$, and the antisymmetry of the pairing on~$U$ means that $Q$ is nilpotent. One expects this space to be isomorphic to~$P^1\times P^7$, corresponding to the kernel and the image. 

Alternatively, one can allow the matrix to have rank two. The product $Q^2$ is then an element of the representation
\deq{
Q^2 \in \wedge^2 S = \rep{7}\oplus\rep{21},
}
where $\rep{7}$ is the vector and $\rep{21}$ the adjoint of~$SO(7)$. The gamma matrices now witness not an isomorphism, but the projection of this onto the~$\rep{7}$; nilpotence then means that the image of this map is entirely in the~$\rep{21}$.

Indeed, one can straightforwardly verify that the variety now has two irreducible components, one of which is the rank-one locus discussed above, and the other of which is the space of pure spinors in ten dimensions. Concretely, 
\deq{
\Nilpp(7,1) \cong \OG(5,10) \sqcup \sigma(P^1 \times P^7)
}
The $\sigma(P^1 \times P^7)$ component contains the images of both of the $P^7$ components in~$\Nilpp(8,1)$ along the dimensional reduction map. The two components of~$\Nilpp(7,1)$ intersect along the space 
\deq[eq:B3A1]{
\OG(3,7) \times P^1  \cong \qu(6) \times P^1 \subset \sigma(P^1 \times P^7).
}
We remind the reader that $\qu(6)$ denotes a generic quadric in~$P^7$; one has that
\deq{
\qu(6) \cong \OG(3,7) \cong \OG(4,8).
}
The commutant $Z^0(Q)$ is easy to compute for a generic~$Q$ on the rank-one component. Indeed, its stabilizer is just the stabilizer in $\Spin(7)$ of a generic element of~$S$, together with the stabilizer in~$\C^2$ of a generic element under the $SU(2)$ $R$-symmetry. The former, essentially by definition, is~$G_2$, and the latter is obviously $U(1)$. So twists of seven-dimensional theories by supercharges in this stratum define theories that make sense on any seven-manifold of~$G_2$ holonomy. This is the weakly topological twist of~\cite{Acharya:1997jn}. 

We remark that the  $n$-th term in the Hilbert series of~\eqref{eq:B3A1} (the coefficient of~$t^n$) is the dimension of the representation of $B_3 \times A_1$ with Dynkin label $[0,0,n;n]$. 
Such behavior is typical for (products of) symmetric spaces.

For the component appearing upon dimensional reduction, the space of exact translations is also easy to compute. The surviving translations will be $W \cap E^\vee$, which is an isotropic subspace of~$\C^7$ of minimal (generic) dimension two, and largest possible dimension three. When three translations survive, we are on the smallest stratum in seven dimensions; when two, we are considering a twist that is topological in three and holomorphic in four (i.e.~two complex) directions.

\section{Extended supersymmetry in six dimensions}

As we already discussed in~\S\ref{ssec:rankstrat}, the essential technique in six dimensions is a generalization of the matrix technique we used in four dimensions. It relies on the fact that the gamma matrices witness isomorphisms
\deq{
V \cong \wedge^2 S_+ \cong \wedge^2 S_-
}
of~$SO(6)$ representations. There is no projection to worry about in this dimension.

\subsection{By dimensional reduction: the $\N=(1,1)$ algebra}

Here, we take $U_+ = U_- = \C^2$, each equipped with the standard symplectic pairing. The odd part of the algebra is thus
\deq{
A^1 = \qty(S_+ \otimes \C^2) \oplus \qty(S_- \otimes \C^2).
}
Importantly, the nilpotence variety now admits a map to~$V\cong\C^5$, obtained by first projecting to~$S_+ \otimes \C^2$, and then taking the pairing. (By nilpotence, projecting to~$S_- \otimes \C^2$ would produce the negative of the same map.) The $R$-symmetry is~$SO(4)$, acting on the space~$U_+\otimes U_-$ via the Dirac spinor representation.

$\Nilpp(6,1,1)$ has two irreducible components, one being just $\OG(5,10)$, and the other being a nine-dimensional variety, which of course contains the $\sigma(P^1\times P^7)$ that appeared in seven dimensions.  The intersection of the two components is some variety of dimension eight. 

The pure spinor space, of course, is smooth, but the nine-dimensional component is singular. Its singular locus is a four-dimensional variety, containing two irreducible components, each in a distinct coordinate $P^7$. (These two $P^7$'s correspond to a purely chiral choice of supercharge.) Each component is just a Segre-embedded $P^1\times P^3$. The singular locus sits inside the intersection of the nine-dimensional component with the pure spinor space.

The nine-dimensional variety is obtained by taking the ideal consisting of the twelve quadrics, six of which define $\sigma(P^1 \times P^3)$ in one coordinate $P^7$ and the other six in the other. The result is not $P^1\times P^3\times P^1 \times P^3$, since only one scaling equivalence is taken. Instead, it is  a fibration over~$P^1$ with generic fiber $P^1\times P^3\times P^1 \times P^3$. At each pole, one of the $P^1 \times P^3$'s shrinks to a point, reproducing the singular locus we calculated above.

In fact, this structure is straightforward to understand: the defining equations~\eqref{eq:6d11mat} for the nilpotence variety were of the form
\deq{
Q_+ \omega Q_+^t + Q_- \omega Q_-^t = 0.
}
The twelve defining equations of the nine-dimensional component are thus simply
\deq{
Q_+ \omega Q_+^t = Q_- \omega Q_-^t = 0.
}

For the component arising from ten dimensions, the space of exact translations is now an isotropic subspace of~$\C^6$ of minimal (generic) dimension one, and largest possible dimension three. When three translations survive, we are on the smallest stratum, a $P^1 \times P^3$, which of course corresponds to a holomorphic twist. When two survive, we are considering a twist that is topological in two and holomorphic in four (i.e.~two complex) directions, and we are on the intersection of the pure-spinor variety with the other, nine-dimensional irreducible component. Finally, when we are in the open (generic) stratum of the pure spinor variety, only one translation survives, and the corresponding twist is topological in four and holomorphic in two (one complex) directions. The large stratum of the other component is a weakly topological twist.

We can sum these considerations up with the following partially ordered set:
\begin{equation}
\begin{tikzcd}[column sep = 0.1 em]
\parbox{1.5cm}{ \centering $\OG(5,10)$ \\ (2,0)} & & \parbox{1cm}{\centering $X^9$ \\ (1,1)}\\
 & \parbox{3cm}{ \centering $\OG(5,10) \cap X^9$  \\ (1,1)} \ar[ru] \ar[lu] & \\
\parbox{1.5cm}{ \centering $P^1 \times P^3$ \\ (1,0)} \ar[ru] & & \ar[lu] 
\parbox{1.5cm}{ \centering $P^1 \times P^3$ \\ (0,1)} 
\end{tikzcd}
\end{equation}
Below each component, we have indicated the ranks of the corresponding supercharge. Note that, in this case, the rank is \emph{not} a complete invariant of the stratification! This is because there is an additional Lorentz invariant: An element of the $(1,1)$ stratum is constructed from one element of~$S_+$ and one element of~$S_-$, and the two spinor representations have a scalar pairing $S_+ \otimes S_- \rightarrow \C$. Requiring this pairing to vanish is a Lorentz-invariant condition; the requirement of $R$-symmetry invariance means that we obtain four extra defining equations in this manner, since the contraction of spinor indices leaves an element of $U_+ \otimes U_-$ (the adjoint of~$U(2)$). Correspondingly, the intersection is generated as a variety by sixteen equations: the twelve defining equations of~$X^9$, together with these four.

\subsection{Chiral supersymmetry: the $\N=(2,0)$ algebra}

To obtain the $\N=(2,0)$ algebra, we take $U_+ = \C^4$ with a standard symplectic pairing $\omega$, and $U_-$ to be empty.
The $R$-symmetry is~$SO(5)$, acting on~$U_+$ via the spin representation. A new feature is that 
\deq{
\wedge^2 U_+ \cong \rep{1} \oplus \rep{5}.
}
After lowering an index with~$\omega$, one can consider $Q$ as a map from~$S^+$ to~$U_+ = \C^4$. The rank of this matrix must be either one or two; when it is one, the supercharge is automatically nilpotent, and we immediately obtain a description of the stratum as~$P^3 \times P^3$. The larger stratum is contained in the rank-two locus of $M^{4\times 4}(\C)$, but strictly. In fact, a rank-two $Q$ is nilpotent if and only if its image is a Lagrangian subspace of~$(U,\omega)$. So (as for any linear map) we can consider the following canonical diagram:
\begin{equation}
\begin{tikzcd}
S_+ \arrow{r}{Q} \arrow[twoheadrightarrow]{d}{\pi} & (U,\omega) \\
S_+ / \ker(Q) \arrow{r}{\sim} & \im(Q) \arrow[hook]{u}{i}
\end{tikzcd}
\end{equation}
We can thus package the data of~$Q$ in terms of its image under the kernel and image maps (which land in Grassmannians), together with the choice of an isomorphism in~$GL(2,\C)$. So we have (for the open stratum) a description
\deq{
PGL(2,\C) \hookrightarrow
\Nilpp(6,2,0)_{\rank=2} \twoheadrightarrow
 \Gr(2,4) \times \Lambda(U,\omega).
}
Here $\Lambda(U,\omega)$ is the Lagrangian Grassmannian.

We can also obtain an algebrogeometric description of this component of the variety. Take the rank-two locus in~$M^{4\times4}(\C)$; there is an image map to subspaces of~$U$. Take any two linearly independent columns $u$ and~$v$ of~$Q$; the Lagrangian constraint on the image amounts to the single quadric $\omega(u,v)=0$.  

On the rank-one locus, we're in the setting of a holomorphic twist; one $P^3$ is the space $\OG(3,6)$, and the other is the projectivization of~$U=\C^4$. On the larger stratum, we obtain a twist that is topological in four and holomorphic in two (one complex) directions~\cite{Gaiotto:2009we, Witten:2011zz}. This can be easily seen as follows: A rank-two supercharge takes the form
\deq{
Q = s\otimes u + t \otimes v \in S_+ \otimes U.
}
With respect to $s$, we can decompose 
\deq{
S_+ = \wedge^0 L \oplus \wedge^2 L,\quad L^\vee = \Ann(s).
}
Now, deforming to a rank-two corresponds to choosing $t \in \wedge^2 L$, and the common annihilator of~$t$ and~$s$ will then be a dimension-one subspace of~$L^\vee$ (spanned by $t \in \wedge^2 L \cong L^\vee$).

As a  pure spinor superfield, the structure sheaf of this nilpotence variety produces the $\N=(2,0)$ tensor multiplet; this was demonstrated in~\cite{Cederwall-coho}. We will review that result briefly in~\S\ref{ssec:CE6d}.

\section{Dimensional reduction to five dimensions: 5d $\N=2$}

The $\N=2$ supersymmetry algebra in five dimensions can be obtained either from dimensional reduction of the $\N=1$ algebra in ten dimensions or the six-dimensional $\N = (2,0)$ supersymmetry algebra.  Consequently, the nilpotence variety $\Nilpp(5,2)$ must contain the union of the two corresponding nilpotence varieties.  In fact, $\Nilpp(10,1)$ and $\Nilpp(6,2,0)$ appear as the two irreducible components of $\Nilpp(5,2)$:
\deq{
\Nilpp(5,2) \cong \Nilpp(10,1) \cup \Nilpp(6,2,0).
}
In five dimensions the nilpotence variety $\Nilpp(5,2)$ is a complete intersection of five quadrics in $P^{15}$ and is therefore of degree $2^5 = 32$.  This setup is essentially the definition of the varieties $\Nilpp(10,1)$ and $\Nilpp(6,2,0)$ being {\it linked} in the sense of liaison theory.  A simple consequence is that the degrees are additive: $\Nilpp(10,1)$ is of degree~$12$ and $\Nilpp(6,2,0)$ is of degree~$20$, so that their sum is $32=2^5$.

A sketch of the stratification is shown below:
\begin{equation}
\begin{tikzcd}[row sep = 1 em, column sep = 1 em]
          \Nilpp(10,1) & & \Nilpp(6,2,0) \\
          &  \Nilpp(10,1) \cap \Nilpp(6,2,0) \arrow[lu] \arrow[ru] & \\
          & \text{sing. }P^3\times P^3\ar[u] & 
\end{tikzcd}
\end{equation}
The intersection of the two irreducible components is an irreducible variety of dimension nine. Its  Hilbert series is
\deq{
\frac{1-11t^2+16t^3+10 t^4 - 32 t^5 + 10 t^6 + 16 t^7 - 11 t^8 + t^{10}}{(1-t)^{16}} 
=
\frac{(1 - t)^6 (1 + t)^2 (1 + 4 t + t^2)}{(1-t)^{16}} 
.
}

Dimensionally reducing the holomorphic twist, one gets precisely two surviving translations, corresponding to a minimal twist in five dimensions. (The image of the dimensional reduction map from the holomorphic locus of~$\Nilpp(6,1,1)$ sits along a $P^1 \sqcup P^1 \subset P^3$.) Upon dimensional reduction of the AGT twist of $\N=(2,0)$, either one or zero translations survive; thus, the corresponding maximal stratum is a purely topological twist. Similarly, dimensional reduction from ten dimensions can leave either two, one, or zero surviving translations.

\section{Maximal supersymmetry in four dimensions}

For $\N=4$ supersymmetry, the whole rank stratification poset appears, so that the stratified decomposition of the variety looks as follows:
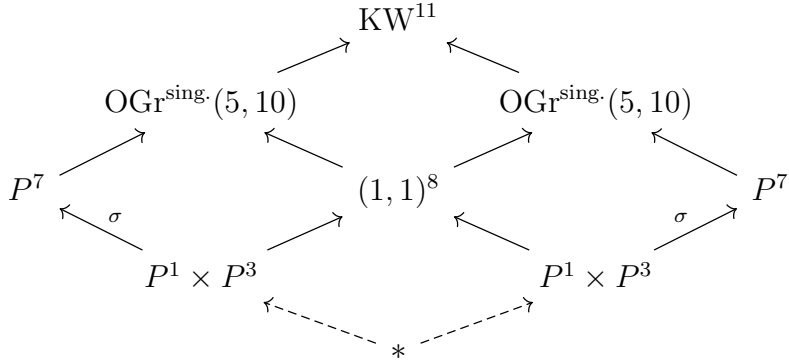
\[
\begin{tikzcd}[row sep=1 em, column sep = 1 em]
  &   &    \text{KW}^{11} & & \\
& \OG^\text{sing.}(5,10) \arrow[ru] &  & \OG^\text{sing.}(5,10) \arrow[lu] & \\
       P^7 \arrow[ru]  &  & (1,1)^8 \arrow[ru] \arrow[lu]  & &  P^7 \arrow[ul]  \\
&  {P^1 \times P^3} \arrow{lu}[swap]{\sigma} \arrow[ru]  &  &  {P^1 \times P^3}  \arrow[lu] \arrow{ru}{\sigma}  & \\
  &   &    * \arrow[dashed,ru] \arrow[dashed,lu] & & \\
\end{tikzcd}
\]
We label the components with superscripts indicating their dimension; for the computation of such dimensions, see~\eqref{eq:4ddims} below. Here $\OG^\text{sing.}(5,10)$ denotes a singular degeneration of the usual smooth orthogonal Grassmannian. The singular locus of that variety is precisely the $P^7$ that is the image of the inclusion of the $(2,0)$ stratum. At the level of the free resolution, the degeneration is witnessed by the appearance of a pair of three-dimensional free modules that cancel at the level of the Hilbert series. Relatedly, while a generic linear section of $\OG(5,10)$ is a canonical genus seven curve~\cite{MR1363081},
the generic linear section of~$\OG^\text{sing.}(5,10)$ is a tetragonal genus seven curve.

The eight-dimensional component, labeled $(1,1)^8$ in the diagram, is the (intrinsic) singular locus of the variety~$\Nilpp(4,4)$. 
It has Hilbert series
\begin{multline}
 \frac{1 - 16 t^2  + 32 t^3  + 14 t^4  - 112 t^5 + 140 t^6  - 64 t^7  - 7t^8  + 16t^9 - 4t^{10}}{(1-t)^{16}}
\\ = \frac{(1 - t)^7 (1 + 2 t) (1 + 5 t + 2 t^2)}{(1-t)^{16}}.
\end{multline}

It is clear to see from explicit representatives that the supercharges produced in twists of either Donaldson--Witten or Kapustin--Witten type are of rank (2,0). Thus, we are in a situation where the twisting map $\phi$ is not uniquely fixed by a choice of supercharge. In the Kapustin--Witten--Marcus twist, supercharges are generically linear combinations of a left-chiral and a right-chiral supercharge of Donaldson type, and the generic rank is therefore (2,2). The twists defined on the  (2,1) loci are only weakly topological.

One can immediately write diagrams, describing pure strata, that are the analogues of those we wrote down for $\N=3$ supersymmetry:
\begin{equation}
\begin{tikzcd}
&U / \ker Q_- \arrow{r}{GL(r_-)} & \im Q_- \arrow[hookrightarrow]{d}{} \\
S_+ \arrow{r}{Q_+} \arrow[twoheadrightarrow]{d}{} & U = \C^4 \arrow[twoheadrightarrow]{u}{} \arrow{r}{Q_-} & S_- .\\
S_+/\ker Q_+ \arrow{r}{GL(r_+)} & \im Q_+  \arrow[hookrightarrow]{u}{} & 
\end{tikzcd}
\end{equation}
From this, one gets the general description of the open stratum as a bundle:
\deq[eq:genstrat]{
P\qty[ GL(r_+) \times GL(r_-) ] \hookrightarrow 
Y_{(r_+,r_-)} \twoheadrightarrow
\Gr(r_+,S_+) \times \Gr(r_-,S_-) \times \Fl(r_+,\N-r_-;U).
}
Recalling the standard dimension formula 
\begin{align}
\dim \Fl(r_+,\N-r_-;U) &= \frac{1}{2} \qty(  \N^2 - r_+^2 - r_-^2 - (\N - r_+ - r_-)^2 ) \\ 
&= \N(r_+ +  r_-) - r_+^2 - r_-^2 - r_+ r_-,
\end{align}
we can then compute a dimension formula for the stratum of the variety:
\begin{align}
\dim Y_{(r_+,r_-)}& = r_+ (2 - r_+) + r_- (2 - r_-) + (r_+^2 + r_-^2 - 1) + \dim \Fl(r_+,\N-r_-;U) 
\\
&=  (\N+2)(r_+ + r_-) - r_+^2 - r_-^2 -  r_+ r_- - 1.
\label{eq:4ddims}
\end{align}
The affine dimension of the algebraic tangent space $T_Q\qty( \Nilp(4,\N))$ is
\deq[eq:daffTQ]{
4(\N - 1) + (2 - r_{+})(2 - r_{-}).
}
We call  $(2 - r_{+})(2 - r_{-})$ the excess dimension. When $\N\geq 2$, it is nonzero precisely when the stratum in question lies in the singular locus of~$\Nilpp(4,\N)$, which is therefore the downward closure of the $(1,1)$ locus. It is easy to check that, for the cone point, the formula~\eqref{eq:daffTQ} returns $4\N$, the dimension of~$A^1$.

The above description~\eqref{eq:genstrat} tells us that the open $(2,2)$ stratum is fibered over~$\Gr(2,4)$ with fiber $P[GL(2)\times GL(2)]$. The fiber admits a map to $P^1$, whose image misses two points: the map is 
\deq{
[m:n] \mapsto
\frac{\det(m)}{\det(n)},
} 
where $m,n \in GL(2)$. Precisely at the poles of this~$P^1$, the rank drops to~$(2,1)$ or~$(1,2)$.
Thus, it is tempting to identify this $P^1$ with the one appearing in \cite{Kapustin-Witten}, and look for a fibration over that space. 

In fact, we can construct the variety $\Nilpp(10,1)$ explicitly as a singular fibration over~$P^1$.
The generic fiber is a copy of the five-dimensional nilpotence variety $\Nilpp(5,2) \cong \Nilpp(10,1) \cup \Nilpp(6,2,0)$.  Over the $P^1$, the $\Nilpp(6,2,0)$ is constant (although it disappears at the singular fibers).  However, the $\OG(5,10)$ varies, and degenerates at the poles to~$\OG^\text{sing.}(5,10)$.  An explicit description is
\begingroup
\allowdisplaybreaks
\begin{align*}
    x_{ 0,0 }y_{ 0,0 }+x_{ 0,1 }y_{ 0,1 }+x_{ 0,2 }y_{ 0,2 }+x_{ 0,3 }y_{ 0,3 } 
 &= 0 \\
     x_{ 0,0 }y_{ 1,0 }+x_{ 0,1 }y_{ 1,1 }+x_{ 0,2 }y_{ 1,2 }+x_{ 0,3 }y_{ 1,3 } 
   &= 0 \\
        x_{ 1,0 }y_{ 0,0 }+x_{ 1,1 }y_{ 0,1 }+x_{ 1,2 }y_{ 0,2 }+x_{ 1,3 }y_{ 0,3 }
    &= 0  \\
   x_{ 1,0 }y_{ 1,0 }+x_{ 1,1 }y_{ 1,1 }+x_{ 1,2 }y_{ 1,2 }+x_{ 1,3 }y_{ 1,3 }
    &= 0  \\
  s(x_ { 1,0 }x_{ 0,1 }-x_{ 0,0 }x_{ 1,1 }) + t (-y_{ 1,2 }y_{ 0,3 }+y_{ 0,2 }y_{ 1,3 }) 
 &= 0  \\
  s(  -x_{ 1,1 }x_{ 0,3 }+x_{ 0,1 }x_{ 1,3 }) + t(-y_{ 1,0 }y_{ 0,2 }+y_{ 0,0 }y_{ 1,2 })
  &= 0\\
  s(   -x_{ 1,0 }x_{ 0,3 }+x_{ 0,0 }x_{ 1,3 })+t(y_{ 1,1 }y_{ 0,2 }-y_{ 0,1 }y_{ 1,2 })
 & = 0 \\
   s(-x_{ 1,1 }x_{ 0,2 }+x_{ 0,1 }x_{ 1,2 })+t(y_{ 1,0 }y_{ 0,3 }-y_{ 0,0 }y_{ 1,3 })
  &  = 0\\
  s(-x_{ 1,0 }x_{ 0,2 }+x_{ 0,0 }x_{ 1,2 })+ t(-y_{ 1,1 }y_{ 0,3 }+y_{ 0,1 }y_{ 1,3 })
      &= 0\\
  s(x_{ 1,2 }x_{ 0,3 }-x_{ 0,2 }x_{ 1,3 }) + t(-y_{ 1,0 }y_{ 0,1 }+y_{ 0,0 }y_{ 1,1})
    & = 0, 
    \end{align*}
    \endgroup
where $[s:t]$ are the projective coordinates of $P^1.$  

Thinking of the data of the supercharges as a pair of two-by-four matrices $x$ and~$y$, it is easy to give an interpretation for the equations appearing above. The first four equations are the four-dimensional nilpotence conditions, which state that the two matrices multiply to zero. The other six equations come from taking a linear combination, determined by $[s:t]$, of a minor of~$x$ with the corresponding conjugate minor of~$y$. The resulting equations are those for the nilpotence variety~$\Nilpp(6,1,1)$.
The map from the global variety can be obtained by letting $s$ be the first minor of the matrix $x$ and $t$ be the complementary matrix $y$, so that the ten equations  above correspond to a splitting of the ten-dimensional spacetime as $10 = 4 + 6$.

\section{Theories with 32 supercharges}

We study these examples  in somewhat less detail than the previous ones, in part because of computational difficulties. The eleven-dimensional nilpotence variety was previously considered in~\cite{Movshev:2011cy,Berkovits:2005hy}; that for IIB supergravity in~\cite{IIB}. We offer a conjectural description of the type IIA nilpotence variety which we believe to be new. 

Both the type~IIA and type~IIB supersymmetry algebras have twists that are topological in four-dimensions and holomorphic in six dimensions.  This is reflected in the classical duality between the topological string partition function on Calabi--Yau threefolds and higher derivative F-terms in the four-dimensional effective action of the compactification of type II string theory on the Calabi--Yau threefold~\cite{Antoniadis:1993ze, Bershadsky:1993cx}.  This can be viewed as a 4d--6d correspondence.
Berkovits considered the twist of the pure-spinor superstring in~\cite{Berkovits-NMPS}.  It would be interesting to perform the twist directly in the spacetime.

Similarly in eleven dimensions, topological $M$-theory~\cite{Dijkgraaf:2004te} can be expected to compute $F$-terms in four-dimensional theory arising from $G_2$~compactifications.   The theory was considered from the point of view of pure spinors in~\cite{Grassi:2004xr}.

\subsection{Eleven dimensions and $G_2$}

In eleven dimensions, the nilpotence variety $\Nilpp(11,1)$ consists of the first two filtered pieces of the Cartan decomposition~\eqref{eq:Cartanstrat} of the minimal eleven dimensional spinor:
\deq{
\PS_0 = \OG(5,11) \subseteq \PS_3 = Y(11,1) \subseteq  \P^{31}.
}
The variety is twenty-two dimensional and its singular locus is the smooth variety $\OG(5,11) \cong \OG(6,12)$; on this locus, the unbroken symmetry algebra is~$Z^0(Q)_\R = \lie{u}(5)$.
  The full stratification was determined by Igusa~\cite{MR0277558}.
The stabilizer of a generic point $Q \in Y(11,1)$ is a parabolic subgroup with Levi factor $G_2 \times \lie{gl}(2)$ \cite{MR0277558, Movshev:2011cy}; in physical language, this corresponds to the holomorphic twist of the four-dimensional $\N=1$ theory resulting from compactification of $M$-theory on a $G_2$ manifold.
The local geometry near the singular locus is described in~\cite{Movshev:2011cy} and~\cite{Berkovits:2005hy}.
The singular locus 
$PS_0 \cong \OG(5,11)$  can be explicitly described by the set of equations
\deq{
Q \Gamma^{\mu} Q = Q \Gamma^{\mu \nu} Q = 0.
}
The degree of the nilpotence variety is 220.

\subsection{Ten dimensions: Type IIA and IIB}

Based on computational evidence, we expect that $\Nilpp(10,1,1)$ is the union 
\deq{
\Nilpp(10,1,1) \cong \Nilpp(11,1) \cup Z.
}
where $Z$ is a complete intersection of 10 quadrics in $P^{31}$.  
The degree of $\Nilpp(10,1,1)$ is 220, the same as the degree of the eleven-dimensional nilpotence variety.  This is consistent with our conjectural description, because the degree of a reducible variety is the sum of the degrees of its top dimensional components. The Hilbert series appears in~\cite{MovshevSchwarzXu}.

The degree of $\Nilpp(10,2,0)$ is 292. The variety was studied in~\cite{IIB}, where the $SO(10)$-equivariant decomposition of its free resolution was calculated. Using the pure-spinor superfield technique, we can identify the list representations in~\cite{IIB} with the BV complex of type IIB supergravity. See~\S\ref{ssec:IIB} for a short exposition of this result; a fuller discussion is relegated to forthcoming work~\cite{twist-modules}.
Closer study of both of these varieties should be related to the conjectural twists of type~II supergravity theories studied in~\cite{CostelloLi}.

\subsection{Nine dimensions and TQFT}

Similar to the five-dimensional nilpotence variety $\Nilpp(5, 2)$, we expect that the nine-dimensional nilpotence variety $\Nilpp(9, 2)$ is the union of the eleven-dimensional variety $\Nilpp(11, 1)$ and the ten-dimensional nilpotence variety $\Nilpp(10, 2,0)$.
The nine-dimensional nilpotence variety $\Nilpp(9, 2)$ is a singular complete intersection of nine quadrics and hence has degree $2^9 = 512.$  This is equal to the sum of the degrees $512 = 220 + 292$ of the eleven-dimensional and ten-dimensional type IIB nilpotence varieties.  By the general arguments of~\S\ref{ssec:dimred}, we expect that there is a (weakly) topological twist in this dimension, arising from the generic dimensional reduction of the larger stratum of~$\Nilpp(11,1)$. It should be definable on the product of a $G_2$ manifold and a (topological) Riemann surface.

\section{The pure spinor superfield technique}
\label{sec:PSSF}

In this section, we discuss the original motivation for the appearance of some of the varieties $\Nilpp(d,\N)$ in the physics literature: namely, the pure spinor superfield formalism. We review the technique in our language, which is perhaps more mathematical than the standard, and point out some analogies with other ideas in the literature, in particular with the theorem of Borel--Weil--Bott. At the end, we will arrive with a clean definition of the pure-spinor superfield technique in terms of reasonably standard mathematical constructions. We begin with a brief discussion of supermanifolds, needed to construct the $A$-module $\Fun(\C^{d|k})$ (an unconstrained superfield) which will be of essential use later. 

As is well-known, a manifold or a space can naturally be represented as some kind of commutative algebra; this can be expressed by saying that there is a contravariant functor from spaces to algebras, consisting of taking functions. This basic theme is expressed repeatedly in different settings all over twentieth-century mathematics; for example, it is at the core of algebraic geometry, but also appears in the theory of $C^*$-algebras (the Gelfand--Naimark theorem characterizes the image of such a functor from locally compact Hausdorff spaces to $C^*$-algebras, upgrading it to an equivalence of categories). 
By an application of the same intuition, a ``supermanifold'' is a space whose algebra of functions is graded by~$\Z/2\Z$ and commutative with respect to the Koszul sign rule. For basic references on supermanifolds and supergeometry, see e.g.~\cite{Bernstein,DeWitt,Berezin}.

The basic supermanifolds are the affine spaces $\C^{d|k}$, where $d$ denotes the even and $k$ the odd ``dimension.'' From an algebrogeometric point of view, affine space is characterized by having a free polynomial algebra of functions:
\deq{
\C^d = \Spec \C[x_1,\ldots,x_d] = \Spec \qty( \Sym^*(\C^d)^\vee ).
}
Similarly, an affine supermanifold has a free $\Z/2$-graded algebra of functions:
\deq{
\Fun(\C^{d|k}) = \C[x_1,\ldots,x_d; \theta_1,\ldots,\theta_k]  \cong \Sym^*(\C^d) \otimes \wedge^*(\C^k).
}
Here, we are content to ignore all analytic issues, and just think of polynomials in a naive way as a basis of some space of functions. 

The super-Poincar\'e algebras we have been writing down act naturally on affine supermanifolds: when $d=\dim V$ and~$k= \dim A^1$, then $A$ is the algebra of affine symmetries of~$\C^{d|k}$. It acts via a set of (super-)translation operators. (We identify $\C^d = V^\vee$ and~$\C^k = (A^1)^\vee$.)
In fact, $\Fun(\C^{d|k})$ is equipped with 
two mutually commuting actions of supertranslations, on the left and on the right. 

The corresponding operators are typically written ${Q}$ and~$\Berk$, and are constructed as follows. Let $s$ be a fixed element of~$A^1$. There are natural ``derivative'' maps
\deq{
\nabla: V \otimes \Sym^{*}(\C^d) \rightarrow \Sym^{*-1}(\C^d), \qquad
A^1 \otimes \wedge^*(\C^k) \rightarrow \wedge^{*-1}(\C^k),
}
which can just be thought of as the interior product.
We can also consider the anticommutator map as defining a map
\begin{equation}
\begin{tikzcd}
\gamma: A^1 \otimes \qty( \Sym^{*}(\C^d)\otimes \wedge^*(\C^k) ) 
\arrow[dashed]{dr}{}
\arrow{r}{\{\cdot,\cdot\}}
&
 V \otimes (A^1)^\vee \otimes \Fun (\C^{d|k}) \arrow{d}{(\nabla,\wedge)}
 \\
 &  \Sym^{*-1}(\C^d)\otimes \wedge^{*+1}(\C^k) ,
 \end{tikzcd}
\end{equation}
In the map at right, $V$ acts by~$\nabla$, and~$A^1$ by the wedge. One then takes $s$ to act on~$\Fun(\C^{d|k})$ by the operator
\deq{
Q_s =  \gamma_s + \nabla_s. 
}
One can then check that
\deq{
\{Q_s,Q_t \} = \nabla_{\{s,t\}},
}
so that this indeed defines a representation of~$A$. 
One could also, however, define the operators
\deq[eq:defDs]{
\Berk_s = i \qty( \gamma_s - \nabla_s ). 
}
And it is not difficult to see that these also define a representation of~$A$ such that the actions of supercharges commute: $\{Q,\Berk\} = 0$. (In both cases, $V$ acts simply by~$\nabla$, and~$\lie{so}(V)$ acts in the obvious way.)

In our language, the chief idea of the pure spinor superfield formalism is as follows. As pointed out in the introduction, any $A$-module defines a family of chain complexes over the (here affine) nilpotence variety $\Nilp(d,\N)$. We can make a more global version of this statement by considering the coordinate ring of the nilpotence variety; the statement is then that, for $M$ an $A$-module, there is a natural chain complex 
\deq[eq:globaltwist]{
\left( M \otimes \OO \qty[ \Nilp(d,\N) ], ~ \Berk = u^\alpha \Berk_\alpha \right) ,
}
where $\Berk_\alpha$ is a basis for~$A^1$ (acting on~$M$) and~$u^\alpha$ is a corresponding set of coordinate functions on~$\Nilp(d,\N)$, which just act by multiplication. This is a global version of the ``twist'' of~$M$ at a particular chosen point~$\Berk \in \Nilp(d,\N)$, which is just the chain complex 
\deq{
\left( M,\,  \Berk \right) .
}
To express $\Berk$ in coordinate-free language, note that the linear coordinate functions are elements of 
\deq[eq:coordring]{
(A^1)^\vee \hookrightarrow \Sym^* \qty( \C^k = (A^1)^\vee ) \twoheadrightarrow  
\Sym^*  \qty( \C^k ) / I  = \OO\qty[ \Nilp(d,\N) ] .
}
They are, in fact, functions on the parity reversal of~$A^1$, which is the affine space in which~$\Nilp(d,\N)$ is defined.
So they live in a space which is dual (as a vector space, ignoring grading and algebraic properties) to $A^1$ itself. There is thus a canonical element 
\deq{
\id \in \End(A^1) = A^1 \otimes (A^1)^\vee.
}
It is important to note, though, that the action of this element will have nothing to do with the algebraic structure on~$\End(A^1)$! We now push forward along the map~\eqref{eq:coordring} to the coordinate ring on the $(A^1)^\vee$ factor, giving us a canonical operator
\deq{
\Berk \in A^1 \otimes \OO\qty[ \Nilp(d,\N) ]
}
which is now nilpotent, since the ideal which is set to zero in~$\OO\qty[ \Nilp(d,\N) ]$ is precisely the ideal where the anticommutator map on the first factor is nonzero. We can now allow (for example) the $A^1$ factor to act by the representation~\eqref{eq:defDs}, and the $\OO\qty[ \Nilp(d,\N) ]$ factor to act by multiplication on itself, recovering~\eqref{eq:globaltwist}.

But this discussion should make it clear that one could even twist~\eqref{eq:globaltwist} by any bundle over~$\Nilp(d,\N)$ that is equivariant with respect to the group action on the variety. To see this, recall that (by the Serre--Swan theorem) such a bundle is precisely a finitely-generated projective module over~$\OO \qty[ \Nilp(d,\N) ]$, after applying the functor of global sections. We constructed a nilpotent operator above by allowing the coordinate functions $u^\alpha$ to act by multiplication; however, they might as well act on any other module. So we could generalize the above and write
\deq[eq:bundles]{
\left( M \otimes \Gamma \qty[ E \rightarrow \Nilp(d,\N) ], ~ \Berk \right) .
}
The requirement of equivariance ensures that~$\Gamma \qty[ E \rightarrow \Nilp(d,\N) ]$ will carry an action of~$A^0$, and therefore that Lorentz and $R$-symmetry will be preserved by the construction.

Similarly, $M$ could \emph{a~priori} be any $A$-module whatsoever. But the module $\Fun(\C^{d|k})$ is special, in that it admits \emph{two} actions of~$A$. So, even after performing the global twist~\eqref{eq:globaltwist} along one action of~$A$, the other action remains unbroken, and we obtain a chain complex of~$A$-modules. In other words, the homology of~\eqref{eq:globaltwist} consists of supermultiplets. So it's immediate to try and start building familiar supermultiplets via global twists of  the unconstrained superfield, of the form
\deq[eq:PSSF]{
\left( \Fun(\C^{d|k}) \otimes \Gamma \qty[ E \rightarrow \Nilp(d,\N) ], ~ \Berk \right) .
}
Then, one can (hope to) understand classification of supermultiplets geometrically, just by studying bundles on nilpotence varieties---analogous to how the Borel--Weil--Bott theorem allows one to construct representations of Lie groups from line bundles on homogeneous spaces. This program was first implemented by Berkovits~\cite{BerkovitsPS,Berkovits-coho}, in the case of ten-dimensional super Yang--Mills theory; in this instance, it is a special case of the pure spinor formalism for the superstring, which is also based on the space~$\Nilp(10,1)$. Generalizations have been pursued by Martin Cederwall and collaborators, among others, most extensively for theories with maximal supersymmetry; see for example~\cite{Cederwall-coho,Cederwall-6d}, as well as~\cite{Movshev:2011cy}. A recent review, including discussion of actions for pure spinor superfields, is~\cite{Cederwall-review}.

In fact, there is no reason that~$\OO\qty[ \Nilp(d,\N) ]$ couldn't act on a space of functions with support just on particular strata of~$\Nilp(d,\N)$, or sections of an equivariant bundle over a stratum! (In other words, it can act on the global sections of any equivariant coherent sheaf.) While these will no longer be precisely the same kinds of modules for~$\OO \qty[ \Nilp(d,\N) ]$, the support conditions respect the action of Lorentz and $R$-symmetry, and can therefore be imposed without any problem. 

We mention also that the \emph{non-minimal} version of the pure spinor formalism~\cite{Berkovits-NMPS} just replaces $\OO \qty[\Nilp(d,\N) ]$ by its resolution, the antiholomorphic de~Rham complex of~$\Nilp(d,\N)$:
\deq{
\OO \qty[\Nilp(d,\N) ] \Rightarrow \qty[ \Omega^{0,*}\qty( \Nilp(d,\N) ), \, \bar\partial ].
}
This allows one to construct an integration pairing on pure spinor superfields, and therefore to write down actions for theories constructed in this way. For more detail, the reader is referred to~\cite{Cederwall-review}.

\subsection{Nilpotence varieties and Chevalley--Eilenberg cohomology}
\label{ssec:CEcoho}

Here, we remark on a simple theorem, which perhaps indicates why the nilpotence variety is of such interest in the study of the super-Poincar\'e algebra itself: it appears in a universal fashion when considering (super) Lie algebra cohomology. While we independently derived these results, they are certainly not new; they appear in some form, for instance, in~\cite{MovshevSchwarzXu}, where Lie algebra cohomology of super-Poincar\'e algebras was considered in detail, as well as probably in earlier literature and/or standard folklore.

Recall that a super Lie algebra structure on a $\Z/2\Z$-graded vector space, which we will denote by 
\deq{ 
\lie{g} = \lie{g}_+ \oplus \lie{g}_-,
}
 is equivalent to a degree-one, $\Z/2\Z$-even nilpotent differential on the free (graded) commutative algebra on its dual, shifted by one:
\deq{
\CE \doteq \Sym^* \qty( \lie{g}_+^\vee [1] \oplus \lie{g}_-^\vee[1] ) \cong
\wedge^* \qty( \lie{g}_+^\vee [1] ) \otimes \Sym^* \qty( \lie{g}_-^\vee[1] ) .
}
The resulting complex is graded by $\Z \oplus \Z/2\Z$, and parity is determined by the sum of the two gradings, modulo two. The differential acts on generators by the dual of the bracket map, and is extended as a derivation to the whole algebra. Nilpotence is equivalent to the (super) Jacobi identity. This complex, the Chevalley--Eilenberg complex, then computes the Lie algebra cohomology of~$\lie{g}$. 
 (For a pedestrian review of some of these ideas, see~\cite{CY4}; a more complete review in the context of physics is~\cite{BBH}, which also reviews related notions such as Koszul complexes.)
 
 We now choose $\lie{g}$ to be the purely ``super-translation'' part of the super-Poincar\'e algebra. That is, we take $\lie{g}_- = A^1$, $\lie{g}_+ = V$, and the only nontrivial bracket map to be the anticommutator. This is clearly a subalgebra of~$A$, admitting an action by the Lorentz algebra $\lie{so}(d)$.
 
In this case, the \emph{a priori} $\Z\oplus \Z/2\Z$ grading lifts to a $\Z\oplus\Z$ grading, since we can remember the degree with respect to  the odd and even generators separately:
\deq[eq:specialCE]{
\CE^{p,q} = \wedge^p \qty( V^\vee ) \otimes \Sym^q \qty[ \qty( A^1 )^\vee ]. 
}
The unrefined Chevalley--Eilenberg degree is then $p+q$, and the parity is simply determined by $p \pmod 2$. The $(p,q)$-degree of the Chevalley--Eilenberg differential is $(-1,2)$; it schematically takes the form
\deq{
d v^i = u^\alpha \Gamma^i_{\alpha\beta} u^\beta,
}
where $v^i$ are coordinates on~$V^\vee$ and~$u^\alpha$ on~$(A^1)^\vee$. Of course, the form of the anticommutator appropriate to the situation can be read off in detail from Table~\ref{tab:spinors}.

Now, note that one factor of the decomposition~\eqref{eq:specialCE} is just the coordinate ring
\deq{
R = \Sym^q \qty[ \qty( A^1 )^\vee ]
}
of the affine space which is (the parity reversal of)~$A^1$. So, if we forget the $q$-grading, the spaces $\CE^{p} = \oplus_q \CE^{p,q}$ define a chain complex of free~$R$-modules, which is of the form
\deq{
\CE^* =  \wedge^* V^\vee \otimes R.
}
Moreover, the differential is generated by extending (uniquely as a derivation) a single map from $V^\vee \rightarrow R$, whose image is the ideal $I \subset R$ defining the nilpotence variety! Thus, $\CE$ is just the Koszul complex~\cite{Koszul} of the quadrics in the defining ideal, over~$R$. Under the assumption that these quadrics form a regular sequence, this Koszul complex may be identified with the free resolution of~$R/I = \OO \qty[\Nilp(d,\N)]$ as an $R$-module. (This was also observed in~\cite{MovshevSchwarzXu}.)

Just to remind the reader of the definition:
\begin{definition}
Let $R$ be a ring, and $\{\rho_i\}$ a finite set of $n$ elements of that ring. The Koszul complex associated to this data is the commutative differential graded $R$-algebra
\deq{
\qty( R[\varepsilon_i] \cong \wedge^* R^n , ~ d:\varepsilon_i \mapsto \rho_i ),
}
where $\varepsilon_i$ are anticommuting generators in degree one, and $d$ is extended as a derivation. Given the additional data of an $R$-module $M$, one may tensor the entire construction over~$R$ with~$M$.
\end{definition}

As is clear from the definition, $\CE$ also has a (graded) commutative algebra structure, which induces a multiplication operation on its cohomology, preserving the grading. This means that each of the spaces
\deq{
H^p(\CE) = \bigoplus_q H^{p,q}(\CE,d)
}
is canonically an $H^0(\CE)$-module. Furthermore, the entire construction is Lorentz and $R$-symmetry equivariant. In the context of the pure spinor superfield formalism, such a module can therefore be interpreted as an equivariant (stratified) bundle on the affine  nilpotence variety. Again, we are being loose about precisely what the appropriate notion of ``stratified bundle'' is, but the example of functions with support on an irreducible component or the downward closure of a stratum should be kept in mind. For explicit computations of these Chevalley--Eilenberg cohomology groups, see~\cite{MovshevSchwarzXu}.

\subsection{Pure spinor formalism for four-dimensional $\N=1$ theories}
\label{ssec:4d1}

In this subsection, we work through the details of a very simple example. (After performing these computations, we found a related analysis of the vector multiplet for minimally supersymmetric Yang--Mills theories in~\cite{Movshev-YM}.) One is interested in a very simple space, $\Nilp(4,1)$, which is just two copies of~$\C^2\subset \C^4$ intersecting at the origin:
\deq{
\OO \qty[\Nilp(4,1) ] = \C[u_\alpha, v_\beta]/(u_\alpha v_\beta),
}
where the ideal consists of four quadrics, $\alpha, \beta \in \{1,2\}$. 
Now, as we emphasized above, to construct interesting supermultiplets, one can choose any equivariant module over this ring. In particular, functions on either irreducible component are a quotient of the ring, and therefore in a natural way a module over it. For example, we could take the module
\deq{
\C[u_\alpha],
}
where the action is the obvious one (i.e., $u_\alpha$ acts by multiplication and~$v_\beta$ by zero). 

Applying the logic of the previous section, we take this module over~$\OO \qty[\Nilp(4,1) ]$, together with the module $\Fun(\C^{4|4})$ over the super-Poincar\'e algebra, and look at the cohomology of the Berkovits operator:
\deq[eq:chiralcpx]{
\qty( \Fun(\C^{4|4}) \otimes \C[u_\alpha], ~ \Berk = u_\alpha \Berk_\alpha ).
}

The computation of this cohomology can be simplified by first considering ``zero-mode cohomology''~\cite{Cederwall-review,IIB}, in which the spacetime variables $x_i$ are set to zero. 
This amounts to recalling that the following facts: First off, as a vector space, the pure-spinor superfield complex consists of $\Fun(\C^{d|k})$---which we could bigrade by the degrees of even and odd generators, just as in the Chevalley--Eilenberg example above---tensored by some module of~$\OO\qty[ \Nilp(d,\N)]$. Secondly, in the definition~\eqref{eq:defDs}, $\Berk_\alpha$ is the sum of two terms, one of which ($\nabla_\alpha$) is independent of the even generators of~$\Fun(\C^{d|k})$, and which have bidegree respectively $(0,-1)$ and~$(-1,+1)$. So we could correspondingly separate $\Berk$ into its two bigraded pieces; computing the cohomology of the first is comparatively simple, since the $\Sym^*(\C^d)$ factor is just along for the ride. A spectral sequence, in which the bigrading is broken to a single grading, then accomplishes the passage to the full $\Berk$ cohomology; the unbroken grading is the standard one in supersymmetry, where even generators carry twice the weight of odd ones. We will see that this spectral sequence is often quite easy to understand: the $E_1$ page (zero-mode cohomology tensored with functions on spacetime) gives the field content of a particular theory or supersymmetry multiplet, while the differential appearing on that page and computing~$E_2$  is identified with the BRST differential acting on those fields.\nfoot{In the following discussion, it is assumed for simplicity that all gauge groups are abelian; for discussion of nonabelian theories, see~\cite{Cederwall-review} and references therein.} Higher differentials are, in the examples we are familiar with, absent for degree reasons. So the complex at page one is (up to a shift by one that restores the correct parity) a BRST or BV complex.

For zero-mode cohomology, the complex~\eqref{eq:chiralcpx} then reduces to
\deq{
\qty( \wedge^* S_+ \otimes \wedge^* S_- \otimes \C[u_\alpha], ~ \Berk = u_\alpha \pdv{}{\theta_\alpha} ),
}
where we have used that $\C^{0|4} \cong S_+ \oplus S_-$, and~$\theta_\alpha$ are coordinates on~$S_+$. Of course, we are implicitly tensoring with $\Sym^* (\C^d)$ everywhere. Since it does not appear in the differential, the factor $\wedge^* S_-$ is also just along for the ride, and it is straightforward to check that the cohomology 
\deq{
H^*\qty( \wedge^* S_+ \otimes \C[u_\alpha], ~ \Berk = u_\alpha \pdv{}{\theta_\alpha} ) \cong \C,
}
by a familiar cancellation of bosonic and fermionic generators.
(Alternatively, one could observe that one is just looking at the Koszul complex of the set of all generators $u_\alpha$ of the polynomial ring.) 

The whole complex is therefore equivalent (as a representation of Lorentz symmetry) to just $\wedge^* S_-$, corresponding to a familiar chiral superfield. For degree reasons, the deformation of the differential can produce nothing new, and the full result is just a multiplet of the form $\wedge^* (S_-) \otimes \Fun(\C^d)$, without additional subtleties. Had we chosen the other irreducible component of~$\Nilp(4,1)$, we would have obtained the chiral multiplet of opposite chirality.  

We can also construct the vector multiplet in a similar fashion, using just the structure sheaf $ \OO \qty[ \Nilp(4,1) ] $. The corresponding complex for zero-mode cohomology is
\deq[eq:4d1VM]{
\qty( \wedge^* S_+ \otimes \wedge^* S_- \otimes \C[u_\alpha,v_\beta]/(u_\alpha v_\beta), ~ \Berk = u_\alpha \pdv{}{\theta_\alpha} + v_\beta \pdv{}{\bar\theta_\beta} ) ,
}
with the factor $\Fun(\C^{4|0})$ implicit. In degree zero, we clearly just obtain a copy of~$\C$: the degree-one component is $\OO\otimes (S_+ \oplus S_-)$, and the differential simply cancels all of the generators of~$\OO$. 

The reader will recognize that~\eqref{eq:4d1VM} is the Koszul complex, over the quotient ring $\OO$, for its set of linear generators. In fact, it is clear that this is true in general: In the spectral sequence we defined above, where the $E_1$ page is zero mode cohomology, the $E_0$ page is the Koszul complex, over the coordinate ring~$\OO \qty[\Nilp(d,\N)]$ of the affine nilpotence variety, of the set of linear generators of the affine ring (functions on~$A^1$) of which~$\OO$ is a quotient. (One could also refer to this more invariantly: the ideal of these generators is the maximal ideal corresponding to the cone point of~$\Nilp(d,\N)$.) The entire object is then tensored with functions on spacetime. Clearly, the whole Koszul complex could be tensored by an $\OO$-module, corresponding to choosing a stratified bundle over the nilpotence variety. The mathematically inclined reader can take this as a concise definition of the pure spinor superfield formalism---however, ignoring the above discussion perhaps obscures why the result admits an unbroken representation of~$A$.

Let us return to our simple example, though. In higher degrees, though, things become more interesting. 
In degree one, the kernel of~$\Berk$ consists of the inverse image of the generating ideal $u_\alpha v_\beta$ under the differential; this is generated by the eight elements
\deq{
u_\alpha \bar\theta_\beta, \quad \theta_\alpha v_\beta,
}
together with the antisymmetric combinations $u_1 \theta_2 - u_2 \theta_1$ and its analogue for the right-chiral variables.

 The degree-two piece, however, looks like
\deq{
\qty[ \wedge^2 S_+ \oplus \wedge^2 S_- \oplus (S_+ \otimes S_-) ] \otimes \OO.
}
The antisymmetric combinations precisely cancel the two corresponding generators in degree one, 
and the image of~$S_+ \otimes S_-$ under the differential will cancel all anti\-sym\-me\-trized expressions of the form 
\deq{
u_\alpha \bar\theta_\beta - \theta_\alpha v_\beta.
}
So the homology in degree one is generated by the four representatives
\deq[eqref:VMgens]{
u_\alpha \bar\theta_\beta + \theta_\alpha v_\beta = (u_\alpha + \theta_\alpha)(v_\beta + \bar\theta_\beta),
}
transforming in the vector representation of~$SO(4)$. We need to check that no higher-order terms in the variables of~$\OO$ survive; after the quotient by the defining ideal, these look like
\deq{
u_\alpha u_\gamma \bar\theta_\beta, \quad \theta_\alpha v_\beta v_\gamma,
}
for a total of twelve generators which must arise from the sixteen generators of~$(S_+ \otimes S_-)$, tensored with linear generators of~$\OO$. And it is easy to check that all generators are cancelled in this way, so that the representatives~\eqref{eqref:VMgens} generate the homology over~$\C$.

By a similar computation, the degree-two cohomology is generated by expressions of the form
\deq{
v_\beta \theta_1\theta_2, \quad u_\alpha \bar\theta_1 \bar\theta_2
}
over~$\C$; the representation of Lorentz symmetry is thus that of a Dirac fermion. 
We leave it to the reader as an exercise to check that the homology in degree three is one-dimensional, corresponding to a standard auxiliary field, and that no higher homology is present.

There is one last twist left to consider: the differential on the $E_1$ page, which contains spacetime derivatives. For degree reasons, it can only act between the lowest homology (a scalar) and the Lorentz vector; by inspecting the form of~\eqref{eq:defDs}, it's easy to see that this differential acts by the formula
\deq{
\Berk' = (u_\alpha \bar\theta_\beta + \theta_\alpha v_\beta) \Gamma^{\alpha\beta}_\mu \partial_\mu,
}
which could be rewritten as 
\deq{
\delta A_\mu = \partial_\mu c.
}
Of course, this is the familiar form of the BRST operator for an abelian gauge multiplet.

\section{Applications} 

In this section, with the aim of further justifying our interest in nilpotence varieties, we briefly mention some applications of the ideas developed in the bulk of this work. The results announced here will be further developed in the forthcoming article~\cite{twist-modules}. We compute the Chevalley--Eilenberg cohomology of the six-dimensional $\N=(2,0)$ supertranslation algebra, and identify the pure-spinor BV complex of the type~IIB nilpotence variety as the field content of the corresponding supergravity theory.

\subsection{Chevalley--Eilenberg cohomology of $\N=(2,0)$ supertranslations}
\label{ssec:CE6d}

We have computed the cohomology of the supertranslation algebra, as defined above in~\S\ref{ssec:CEcoho}. The cohomology is displayed in table~\ref{tab:6dCE} below; the Hilbert series are, of course, the graded dimensions of the respective $\OO$-modules. By our general result above, the degree-zero cohomology is just $\OO$; the degree-one cohomology is the global sections of the dualizing sheaf of~$\OO$, which can be identified with the tautological bundle $\Ann$ whose fiber over~$Q$ is the subspace $\Ann(Q) \subset V$---generically of dimension one in this instance.
\begin{table}[ht]
\begin{tabular}{|c|c|l|}
\hline
\textit{Bundle:} & \textit{Multiplet:} & \textit{Hilbert Numerator:} \\
\hline
\hline
$ \OO$ & tensor multiplet & $1 + 5t + 9t^2 + t^3$ \\
$ \Ann$ & dualizing multiplet & $5 + 9t + 5t^2 + t^3$ \\
\hline
\end{tabular}
\label{tab:6dCE}
\caption{Chevalley--Eilenberg cohomology of supertranslations in six-dimensional $\N=(2,0)$. In each case, the denominator of the Hilbert series is~$(1-t)^{11}$.}
\end{table}

When one applies the pure-spinor superfield technique to these two modules, the structure sheaf can be identified with the abelian tensor multiplet~\cite{Cederwall-coho}. The $E_1$ page can be computed explicitly; the result is
\begin{equation}
\begin{bmatrix}
1 & - & - & - & - & - \\
- & 6 & - & - & - & - \\ \hline
- & - & 20 & 16 & - & - \\ \hline
- & - & - & 10 & 16 & 5 
\end{bmatrix} = 
\begin{bmatrix}
1 & - & - & - & - & - \\
- & 6 & - & - & - & - \\ \hline
- & - & 15 & - & - & - \\ \hline
- & - & - & 10 & - & - 
\end{bmatrix}
+
\begin{bmatrix}
- & - & - & - & - & - \\
- & - & - & - & - & - \\ \hline
- & - & 5 & 16 & - & - \\ \hline
- & - & - & - & 16 & 5 
\end{bmatrix}.
\label{eq:20tensor}
\end{equation}
Here, the vertical grading is ghost number, while the horizontal grading determines Grassmann parity. 
Note that the first of the pieces in~\eqref{eq:20tensor} looks like the following complex:
\deq{
\Omega^0(\R^6) \rightarrow
\Omega^1(\R^6) \rightarrow
\Omega^2(\R^6) \rightarrow
\Omega^3_-(\R^6)
}
This is the BV complex of a two-form gauge field with self-dual field strength; the extra antifield is present in order to enforce the self-duality constraint. 
The second piece of~\eqref{eq:20tensor} contains the five scalars and two chiral spinors of the $\N=(2,0)$ tensor multiplet, together with their antifields.

In the dualizing module, the result of the computation is exactly the same, except that fields and antifields have been reversed. This leaves the matter content invariant, but flips the complex associated to the self-dual two-form to
\deq{
\Omega^3_+(\R^6)\rightarrow
\Omega^4(\R^6)\rightarrow
\Omega^5(\R^6) \rightarrow
\Omega^6(\R^6).
}
Overall, the result for the $E_1$ page is
\begin{equation}
\begin{bmatrix}
5 & 16 & 10 & - & - & - \\
- & - & 16 & 20 & - & - \\ 
- & - & - & - & 6 & - \\ 
- & - & - & - & - & 1 
\end{bmatrix} = 
\begin{bmatrix}
- & - & 10 & - & - & - \\
- & - & - & 15 & - & - \\ 
- & - & - & - & 6 & - \\ 
- & - & - & - & - & 1 
\end{bmatrix}
+
\begin{bmatrix}
5 & 16 &  & - & - & - \\
- & - & 16 & 5 & - & - \\ 
- & - & - & - & - & - \\ 
- & - & - & - & - & -
\end{bmatrix}.
\label{eq:20tensor}
\end{equation}
It would be interesting to explore the possiblity of writing an action of $BF$ type, pairing the structure sheaf with its dualizing sheaf.

\subsection{IIB supergravity theory}
\label{ssec:IIB}

As a perhaps somewhat more spectacular combination of ideas from the last two sections, we conjecture a close connection between the IIB nilpotence variety and type~IIB supergravity.
The zero-mode cohomology of the IIB nilpotence variety can be computed with computer algebra software, most easily by considering the free resolution of the structure sheaf as a module over the polynomial ring. The result is as follows:
\begin{equation}
\begin{bmatrix}
1 &  -   &    - & - & - & - & - & - & - & -\\
-  & 10 &   - & - & - & - & - & - & - & -\\
-  &  -   &   47 & - & - & - & - & - & - & -\\
-  &  -   &  - & 150 & 32 & - & - & - & - & -\\ \hline
-  &  -   &  - & -      & 357 & 352 & - & - & - & -\\ \hline
-  &  -   &  - &      - & -      & 126      & 352 & 147 & - & - \\
-  &  -   &    - & - & - & - & - & 32 & 30 & -\\
-  &  -   &    - & - & - & - & - & - & - & 2\\
\end{bmatrix}
\end{equation}
Here, the numbers are the dimensions of the cohomology groups. In the context of the pure-spinor superfield formalism, the vertical grading becomes the ghost number or BV degree, and the horizontal grading determines the Grassmann parity. The BRST differential, as discussed above in~\S\ref{sec:PSSF}, will act diagonally, and appears as the differential on the $E_1$ page; the typical example to keep in mind is 
\[
\delta A_\mu = \partial_\mu c, 
\]
where $c$ is the ghost. Thus the BRST differential has ghost number $+1$.
The physical field content appears in the fifth row of the table; we indicate this with horizontal lines.

This cohomology and its equivariant decomposition were computed in~\cite{IIB}; the decomposition into irreducibles of~$SO(10)$ is reproduced below:
\begingroup
\allowdisplaybreaks[1]
\begin{multline}
\left[ \begin{matrix}
V_0 &  -   &    - & - & - & -  \\
-  & V_{\omega_1} &   - & - & -   &  - \\
-  &  -   &   V_{\omega_2} + 2V_0 & - & - &  -    \\
-  &  -   &  - & V_{\omega_3} + 3 V_{\omega_1} & 2 V_{\omega_5}  &  -  \\ \hline
-  &  -   &  - & -      & 3V_{0} + V_{\omega_4 + \omega_5} + 2 V_{\omega_2} + V_{2 \omega_1} &  4 V_{\omega_4} + 2 V_{\omega_1 + \omega_5}  \\ \hline
-  &  -   &  - &      - & -  &   V_{2 \omega_4}        \\
- & - & - & - & - &  -  \\
- & - & - & - & - &  -  \\
\end{matrix} \right.~  \cdots
\\
\cdots ~\left.
\begin{matrix}
 - & - & - & -  \\
 - & - & - & -  \\
- & - & - & -  \\
 - & -  & - & -  \\ \hline
 - & - & - & -  \\ \hline
  4 V_{\omega_5} + 2  V_{\omega_1 + \omega_4} & 3V_0 + 2 V_{\omega_2} + V_{2 \omega_1} & - & - \\
 - & 2 V_{\omega_4} & 3 V_{\omega_1} & -  \\
 - & - & - & 2 V_0  \\
 \end{matrix} \right]
 \label{eq:IIBreps}
 \end{multline}
 \endgroup
 Here, $V_\omega$ denotes the representation with the corresponding Dynkin label; the $\omega_i$ are simple roots, so that $V_{\omega_1} = [1,0,0,0,0]$ is the vector, $V_{\omega_5} = [0,0,0,0,1]$ a chiral spinor representation, and so on. We have divided the matrix for typographical reasons, but the splitting also corresponds to the division into field and antifield multiplets, as will become apparent shortly.

We can divide the $E_1$ page (the off-shell field content) into the following complexes: Firstly, there is a piece of the form
\begin{equation}
\begin{tikzcd}[column sep = 7 pt, row sep = 7 pt]
\Omega^0(\R^{10}) \ar[dr] & & & & & \\
& \Omega^1(\R^{10}) \ar[dr]& & & & \\
& & \Omega^2(\R^{10}) \ar[dr]& & & \\
& & & \Omega^3(\R^{10}) \ar[dr]& & \\
& & & & \physbox{\Omega^4(\R^{10}) }\ar[dr]& \\
& & & & & \Omega^5_-(\R^{10}).
\end{tikzcd}
\end{equation}
Here, we have boxed the physical  component in ghost number zero, which is a four-form gauge field. It appears together with its complete higher-order ghost system. For this identification, one must remember that 
\deq{
\wedge^4 V_{\omega_1} = V_{\omega_4 + \omega_5}, \quad
\wedge^5 V_{\omega_1} = V_{2 \omega_4}  \oplus V_{2 \omega_5},
}
where the latter decomposition corresponds to the self-dual and anti-self-dual parts of the five-form.
Note than an extra antifield appears; since antifields impose equations of motion or constraint equations, this can be interpreted as imposing the self-duality condition on the field strength of the physical four-form gauge field. Since the complex already contains equal numbers of off-shell bosonic and fermionic degrees of freedom, no antifield complex must appear. A similar self-dual complex appears in the tensor multiplet of the~$\N=(2,0)$ theory~\cite{Cederwall-coho}.

Next, there are two copies of a complex of the following form:
\begin{equation}
\begin{tikzcd}[column sep = 7 pt, row sep = 7 pt]
\Omega^0(\R^{10}) \ar[dr] & & \\
& \Omega^1(\R^{10}) \ar[dr] & \\
& & \physbox{\Omega^2(\R^{10})}\, . 
\end{tikzcd}
\end{equation}
This represents an unconstrained two-form gauge field, together with its system of ghosts. These match the two two-form fields (one in the NS-NS and one in the R-R sector) in type~IIB supergravity. This complex (and in fact all  complexes except the self-dual four-form) will appear together with its set of BV antifields. 

Moving on, there is a complex that looks like
\begin{equation}
\begin{tikzcd}[column sep = 7 pt, row sep = 7 pt]
T(\R^{10}) \ar[dr] & \\
& \physbox{S^2(\R^{10})}\,,
\end{tikzcd}
\end{equation}
where $S^2$ refers to a symmetric two-index tensor. This corresponds to the graviton, together with ghosts for its gauge invariance (corresponding to vector fields generating diffeomorphisms). Note that there are no ghosts for ghosts in this case. Also, one should remember that at the level of~$SO(10)$ representations,
\deq{
\Sym^2 V_{\omega_1} =  V_0 \oplus V_{2\omega_1} ,
}
where the trivial representation is of course the (pure gauge) trace part. Finally, there are two additional scalars, which just look like
\deq{
\physbox{V_0}\,,
}
and correspond to the dilaton and the Ramond-Ramond 0-form. 

These are all of the bosonic degrees of freedom. As for the fermionic physical fields (which are of course of odd parity, and therefore appear in~\eqref{eq:IIBreps} shifted to the right by one), there are two copies of a complex of the form
\begin{equation}
\begin{tikzcd}[column sep = 7 pt, row sep = 7 pt]
S_+(\R^{10}) \ar[dr] & \\
& \physbox{T(\R^{10}) \otimes S_+(\R^{10})}\,,
\end{tikzcd}
\end{equation}
Of course, we mean a spin-$3/2$ field with one vector and one spinor index, in the tensor product of the tangent bundle with the appropriate spinor bundle. At the level of Lorentz representations, one must remember that 
\deq{
V_{\omega_1} \otimes V_{\omega_5} = V_{\omega_4} \oplus V_{\omega_1 + \omega_5},
}
so that the physical content of this subcomplex is a reducible representation of Lorentz. These correspond to the two gravitinos, together with their fermionic gauge invariances. 

Finally, this leaves only two remaining simple complexes of the form
\begin{equation}
\physbox{S_-(\R^{10})}\, ,
\end{equation}
which are obviously to be interpreted as the two dilatinos. 
Of course, in this brief exposition, we have neither explicitly computed the pure-spinor differential, nor showed in detail that it agrees with the standard gauge invariances of the fields of type~IIB. We look forward to reporting more thoroughly on this identification in forthcoming work~\cite{twist-modules}.

\section*{Outlook}

In this work, we have tried to call attention to nilpotence varieties as natural objects of study, and to emphasize that they appear naturally in three \emph{a priori} different contexts of interest in the study of supersymmetric field theories: as the natural moduli spaces parameterizing all possible twists; as crucial ingredients in the pure-spinor superfield formalism, which is a global version of a twisting construction applied to a free superfield; and in the study of the Chevalley--Eilenberg cohomology of the super-Poincar\'e algebra. We have also indicated a couple of applications of these ideas.

There are naturally many open questions and promising directions for future work, and we list just a few of them here. First off, it would be interesting to construct other well-known multiplets in terms of the geometric data of a stratified bundle on the nilpotence variety, to give a general algorithm for constructing such bundles (an inverse construction to the theorems of~\S\ref{sec:PSSF}), or to prove a sharp analogue of Borel--Weil--Bott, demonstrating that all representations of super-Poincar\'e admit such a formulation. Many constructions of multiplets have been given in the literature; see just for example~\cite{Cederwall-6d,Movshev:2011cy}. As briefly described in this note, the same construction for the structure sheaf of the IIB nilpotence variety reproduces the field content of the supergravity theory, and presumably its complete BRST/BV complex.
We will study this in more detail in~\cite{twist-modules}, in which we also hope to report on additional work in this direction.

It also seems promising to study the BV complex constructed in~\S\ref{ssec:IIB} to answer other questions about supergravity. For example, since the action of supersymmetry is known on all fields (including ghost sectors) and the untwisted differential can be computed, this complex could then be directly twisted, using a standard prescription. This might allow one to come to a better understanding of the conjectures of Costello and Li~\cite{CostelloLi}, who proposed a relation between a target-space twist of supergravity and Kodaira--Spencer theory. We hope to report progress in this direction soon.

Furthermore, it seems worthwhile to engage in broader study of twists of theories that are already formulated in pure-spinor superfield language. Apart from the appearance of the nilpotence variety in both contexts, several intriguing coincidences point to an interesting relation between action principles for pure spinor superfields and for twisted theories. For ten-dimensional super Yang--Mills theory, the pure-spinor superfield action is of Chern--Simons type~\cite[for example]{Cederwall-review}.
As was first computed in~\cite{Baulieu}, the holomorphic twist is also of Chern--Simons type (see~\cite{CostelloLi} and~\cite{CY4} for later work related to holomorphic twists of this theory). It would be interesting to construct the spectral sequence relating the free superfield to the complex~$(\Omega^{0,*}(\R^{10}),\bar\partial)$, and to understand why the action functional takes the same form at every step.

It would also be interesting to extend these techniques, whether for analyzing possible twists or for constructing supermultiplets, to the case of superconformal algebras. 
In this case, the relevant super Lie algebras are semisimple, so that the results of Gruson~\cite{Gruson} should directly apply. 
The nilpotence variety of the four-dimensional $\N=2$ superconformal algebra has been computed explicitly, and will be studied in detail in forthcoming work~\cite{GNPS-forthcoming} in the context of superconformal indices.

Lastly, as we have mentioned, it seems reasonable to expect a connection to recent work~\cite{DeAzcarragaTownsend,FiorenzaSatiSchreiber,HuertaSchreiber} constructing brane spectra in string and $M$-theories from an analysis of Chevalley--Eilenberg cohomology of algebras closely related to super-Poincar\'e. We plan to  investigate these last points in the future, and hope that others are motivated to do so as well.

\subsection*{Acknowledgements}

We are grateful to  
S.~Gukov, O.~Gwilliam, S.~Nawata, D.~Pei, and B.~Williams for conversations.  We made heavy use of the computer algebra package \emph{Macaulay2} in the course of our work. I.A.S.\ thanks the Max-Planck-Institut f{\"u}r Mathematik and the University of Bristol for hospitality.

Our work is supported in part by the Deutsche Forschungsgemeinschaft, within the
framework of the Exzellenzinitiative an der Universit{\"a}t Heidelberg.

\bibliographystyle{ytphys}
\bibliography{twist-varieties}
\end{document}